\documentclass{article}
\usepackage[paperwidth=7in, paperheight=10in, margin=.875in]{geometry}
\usepackage{amsfonts,amssymb}
\usepackage{amsmath}
\usepackage{graphicx}
\usepackage{cite}
\usepackage{enumerate}
\usepackage[backref,colorlinks,linkcolor=red,anchorcolor=green,citecolor=blue]{hyperref}
\providecommand{\keywords}[1]{\textbf{{Keywords.}} #1}
\providecommand{\AMS}[1]{\textbf{{Mathematics Subject Classification.}} #1}

   \allowdisplaybreaks
\begin{document}
\title{Asymmetric behavior of surface waves \\
induced by an underlying interfacial wave\thanks{Received date, and accepted date (The correct dates will be entered by the editor).}}


\author{Shixiao W. Jiang\thanks{Department of Mathematics, the Pennsylvania State University, University Park, PA 16802-6400, USA, (suj235@psu.edu). }
\and Gregor Kova\v{c}i\v{c}\thanks{Mathematical Sciences Department, Rensselaer Polytechnic Institute, 110 8th Street, Troy, New York 12180, USA, (kovacg@rpi.edu).}
\and Douglas Zhou\thanks{School of Mathematical Sciences, MOE-LSC, and Institute of Natural Sciences, Shanghai Jiao Tong University, Shanghai 200240, China, (zdz@sjtu.edu.cn).}}

         \pagestyle{myheadings} \markboth{\textsc{Asymmetric behavior of SWs induced by an underlying IW}}{\textsc{S.W. Jiang, G. Kova\v{c}i\v{c}, and D. Zhou}} \maketitle

\begin{abstract}
We develop a weakly nonlinear model to study the spatiotemporal
manifestation and the dynamical behavior of surface waves in the presence of an underlying
interfacial solitary wave in a two-layer fluid system.
We show that interfacial solitary-wave solutions of this model can capture the ubiquitous broadening of large-amplitude internal
waves in the ocean.
In addition, the model is capable of capturing three asymmetric behaviors of
surface waves: (i) Surface waves become short in wavelength at the
leading edge and long  at the trailing edge of an underlying interfacial solitary
wave. (ii) Surface waves propagate towards the trailing edge with a relatively small group velocity,
and towards the leading edge with a relatively large group velocity.
(iii) Surface waves become  high in amplitude at the
leading edge and  low at the trailing edge.
These asymmetric behaviors  can be well quantified in the theoretical framework of ray-based theories.
Our model is relatively easily tractable both theoretically and numerically, thus facilitating the understanding of the surface signature of the observed internal waves.
\end{abstract}



\AMS{76B55; 76B07; 35L05; 65M22}


\section{Introduction}\label{intro}

Internal waves  with large amplitudes and long wavelengths are widely observed in
coastal ocean regions,  and are believed to
be important for transferring momentum, heat, and energy in the ocean \cite{Jackson2004Atlas,Helfrich2006ARFM}. Because of their strong turbulent mixing and
breaking, they can influence many ocean processes, such as nutrient supply,
sediment and pollutant transport, acoustic transmission, and interaction
with man-made structures
\cite{Duda1998DTIC,Alford2015Nature}. Generated by tidal flow, internal waves
usually can propagate thousands of kilometers from their source before
dissipation, sloping, and breaking extinguish them \cite{Helfrich2006ARFM,Apel2007JASA}.
In recent decades,
observation data for internal waves and their corresponding surface signature have been
recorded using \textit{in situ} measurements and Synthetic Aperture
Radar (SAR) in many coastal seas worldwide \cite{Osborne1980Sci,Duda1998DTIC,Alford2015Nature}.

Many previous studies have investigated the interaction between interfacial waves (IWs) and surface waves (SWs)
in two-layer fluid systems.
These works can be broadly subdivided into two classes.
The first class focuses on the issue of addressing the direct numerical simulation of
the Laplace equations for velocity potentials
with the appropriate boundary conditions in a two-layer fluid system \cite{Alam2009JFM,Tanaka2015JFM}.
The system can be formulated as a Hamiltonian system for a set of canonical variables \cite{Ambrosi2000WM,Alam2009JFM}.
To evaluate the time derivatives of these canonical variables,
the high-order spectral (HOS) method \cite{Alam2009JFM} is employed to solve
the boundary value problems of Laplace equations. By applying the HOS method,
it was found that energy can be transferred from
SWs to IWs in the two-layer
fluid system \cite{Tanaka2015JFM}. However, the direct numerical simulations are
typically expensive, and it is also challenging to extract the mechanism underlying their results.

The second class focuses on the reduced models in the two-layer fluid system.
To overcome the difficulty stemming from expensive computations, a common approach is to study a reduced model of the
two-layer Euler system via multi-scale analysis \cite{Kawahara1975JPSJ,Hashizume1980JPSJp,Funakoshi1983JPSJp,Donato1999JFM,Parau2001JFM,Bakhanov2002JGR,Barros2007SAM,Hwung2009JFM,Craig2012JFM}.
These reduced models describing two-layer fluids mostly focus on three aspects:

[a] \textit{Traveling wave solutions}: Traveling-wave
solutions exhibiting oscillations are found in the reduced models.
For example, generalized solitary waves with non-decaying oscillations along their tails in addition to the
solitary pulse were found in a long-wave model
\cite{Dias2001PhysD,Parau2001JFM,Fochesato2005PhysD}.
Besides the generalized solitary waves, multi-humped solitary waves with a finite number of oscillations riding on the solitary pulse were found in a fully-nonlinear long-wave model \cite{Barros2007SAM}.

[b] \textit{Ray-based theories}: Many ray-based studies take a
statistical viewpoint of SWs modulated by a near-surface current induced by
IWs \cite{Gargettt1972JFM,Lewis1974JFM,Caponi1988DTIC,Bakhanov2002JGR,Apel2007JASA}.
These studies invoke phase-averaged models based on a wave-balance
equation and ray-based theory \cite{Caponi1988DTIC,Bakhanov2002JGR}, which can be
applied to the remote-sensing observations of IWs via their surface
manifestations.

[c] \textit{Resonant excitations}: When two different modes coexist in a fluid system, a resonant interaction
becomes possible between the modes to aid in
transferring energy in the ocean \cite{Phillips1974ARFM}.
Class 3 triad resonance  is regarded as being responsible for the surface
signature of the underlying IWs \cite%
{Phillips1974ARFM,Osborne1980Sci,Hashizume1980JPSJp,Funakoshi1983JPSJp,Craig2011NatHazd,Alam2012JFM}.
Based on a class 3 triad resonance condition,
many reduced models have been derived for the interfacial and surface waves%
\cite{Lewis1974JFM,Kawahara1975JPSJ,Hashizume1980JPSJp,Funakoshi1983JPSJp,Sepulveda1987PhysFld,Lee2007JKrPS,Hwung2009JFM,Craig2012JFM}%
.
A detailed discussion of surface signature phenomena of IWs was presented in \cite{Craig2011NatHazd,Craig2004CRM,Craig2005CPAM,Craig2012JFM} using
a coupled Korteweg--de Vries (KdV) and a linear Schr\"{o}dinger model.
Narrow rough regions containing surface ripples were found and interpreted as a result of the energy accumulation
in the localized bound states of the Schr\"{o}dinger equation \cite{Craig2012JFM}.

Different from previous works, we here develop a new reduced model to investigate the spatiotemporal manifestation of the small-amplitude SWs in the presence of an interfacial solitary wave in the two-layer setting. Based on model simulations, we demonstrate that our model is successful in characterizing many types of dynamical behavior of SWs, which can be well understood using the ray-based theories.
In Sec. \ref%
{sec:one_dim_isw}, we derive a reduced model for the two-layer fluid system.
In Sec. \ref{sec:basic_properties}, we analyze the
basic properties of the model, including interfacial solitary-wave solutions
and  dispersion relations.
In Sec. \ref{sec:scheme}, we present the numerical scheme and examine its numerical convergence.
In Sec. \ref{sec:surface_wave}, we show the numerical results for the asymmetric behavior of
SWs and quantify this asymmetric behavior using the ray-based theories.
Conclusions and discussion are given in
Sec. \ref{sec:conclusion}, and some mathematical details are presented in the
Appendix.

\section{The two-layer weakly nonlinear (TWN) model}

\label{sec:one_dim_isw}

%

\begin{figure}
\centering 
\hspace*{-2.0mm} \includegraphics[width=0.70\textwidth,height=0.33%
\textheight]{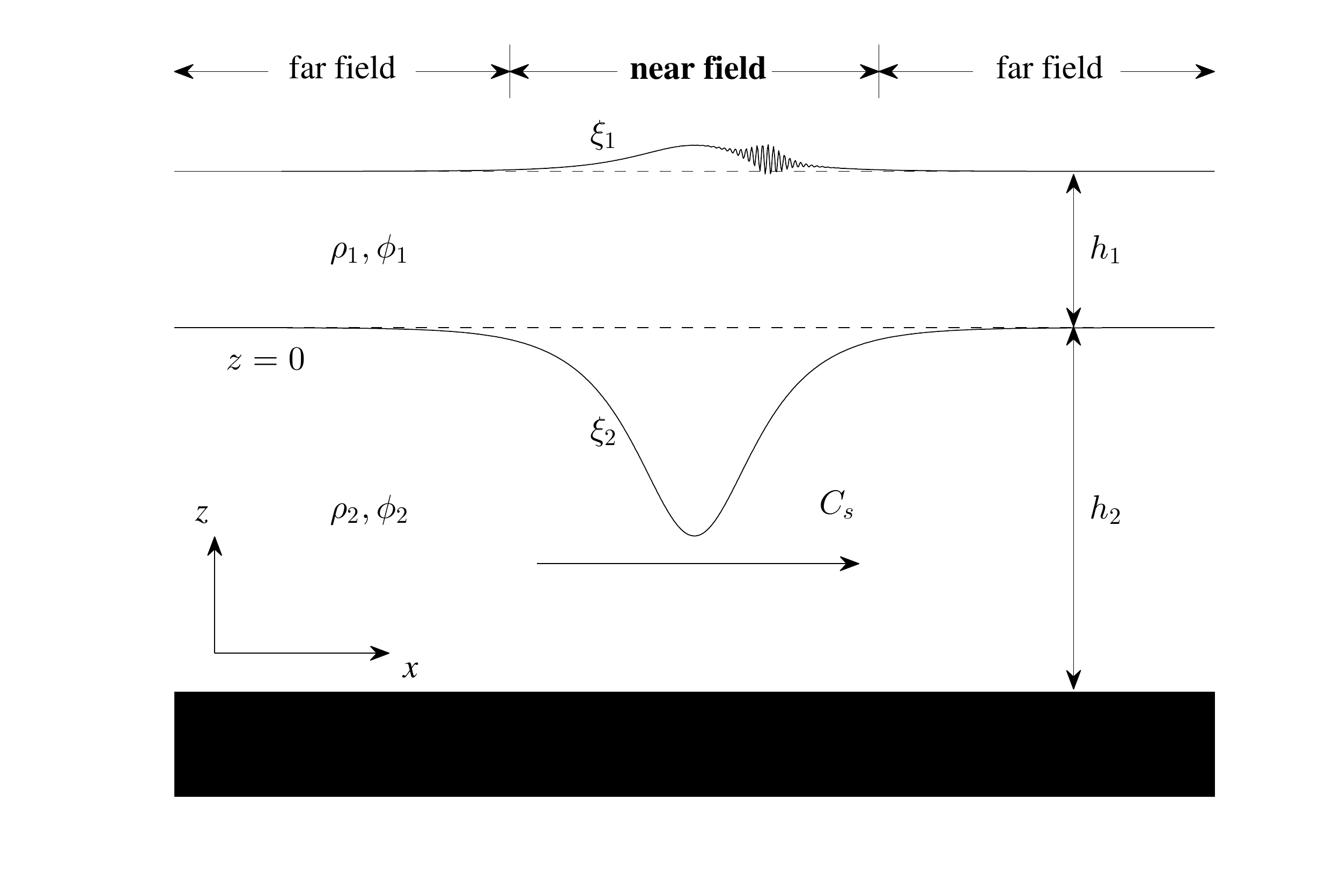}
\caption{ Sketch of the two-layer fluid system  [see text]. }
\label{Fig:sketch_two_layer}
\end{figure}

We first introduce Euler equations for two immiscible
layers of potential fluids with unequal densities.
The two layers of fluids  are assumed to be inviscid, irrotational, and
incompressible. The unequal densities for the upper layer and for the lower layer are denoted by $\rho _{1}$
and $\rho _{2}$, respectively. Here, $\rho _{2}>\rho _{1}$ is assumed for the stable case.
The horizontal and vertical coordinates are $x$ and $z$,
respectively. We focus on the evolution of large-amplitude
interfacial waves $\xi _{2}$ between the two fluid layers, and their coupling with the
the overlaying free surface, $\xi _{1}$ [see Fig.
\ref{Fig:sketch_two_layer}]. The velocity potential $\phi _{i}$ ($i=1$ for the
upper layer and $i=2$\ for the lower layer) satisfies Laplace's equation,
\begin{equation}
\phi _{ixx}+\phi _{izz}=0.  \label{Eqn:1.Laplace}
\end{equation}%
The kinematic equations for the continuity of the normal velocity at the
surface $h_{1}+\xi _{1}$, the interface $\xi _{2}$, and the flat topography $-h_{2}$ are given
in the form
\begin{equation}
\xi _{1t}+\phi _{1x}\xi _{1x}=\phi _{1z},\text{ \ \ at }z=h_{1}+\xi _{1},
\label{Eqn:1.kin_sf}
\end{equation}%
\begin{equation}
\xi _{2t}+\phi _{1x}\xi _{2x}=\phi _{1z},\text{ \ \ at }z=\xi _{2},
\label{Eqn:1.kin_if1}
\end{equation}%
\begin{equation}
\xi _{2t}+\phi _{2x}\xi _{2x}=\phi _{2z},\text{ \ \ at }z=\xi _{2},
\label{Eqn:1.kin_if2}
\end{equation}%
\begin{equation}
\phi _{2z}=0,\text{ \ \ at }z=-h_{2},  \label{Eqn:1.kin_bm}
\end{equation}%
where $h_{1}$ and $h_{2}$ are the undisturbed thicknesses of the upper and lower
layers, respectively. The dynamical equations for the
continuity of pressure at the surface and the interface are the Bernoulli equations,
\begin{equation}
\phi _{1t}+\frac{1}{2}\left( \phi _{1x}^{2}+\phi _{1z}^{2}\right) +g\xi
_{1}=0,\text{ \ \ at }z=h_{1}+\xi _{1},  \label{Eqn:1.dyn_sf}
\end{equation}%
\begin{align}
\rho _{1}\left( \phi _{1t}+\frac{1}{2}\left( \phi _{1x}^{2}+\phi
_{1z}^{2}\right) +g\xi _{2}\right) & =\rho _{2}\left( \phi _{2t}+\frac{1}{2}%
\left( \phi _{2x}^{2}+\phi _{2z}^{2}\right) +g\xi _{2}\right) ,
\label{Eqn:1.dyn_if} \\
\text{at \ \ }z& =\xi _{2}.  \notag
\end{align}%
where $g$ is the gravitational acceleration.

%

For the small-amplitude approximation, we assume that the characteristic
amplitude, $a$, of the IWs and SWs is much smaller than the thickness of the two fluid layers,
\begin{equation}
a/h_{1}=\alpha \ll 1,\ \ \ h_{2}/h_{1}=O(1).  \label{Eqn:1.assm_alpha}
\end{equation}%
For the long-wave approximation, we assume that the thickness of each fluid
layer is much smaller than the characteristic wavelength, $L$, of the IWs and SWs,%
\begin{equation}
h_{1}^{2}/L^{2}=\beta \ll 1,  \ \ \  h_{2}/h_{1}=O(1).\label{Eqn:1.assm_beta}
\end{equation}%
The two small parameters, $\alpha $ and $\beta $, control the nonlinear and
dispersive effects, respectively. Based on the scaling (\ref%
{Eqn:1.assm_alpha})-(\ref{Eqn:1.assm_beta}), we may nondimensionalize all
the physical variables by taking the original variables to be%
\begin{equation*}
x=Lx^{\ast },\text{ \ \ }z=h_{1}z^{\ast },\text{ \ \ }t=(L/U_{0})t^{\ast },
\end{equation*}%
\begin{equation}
\left(\phi _{1},\phi _{2}\right)=\left( aLU_{0}/h_{1}\right)
\left(\phi _{1}^{\ast },\phi _{2}^{\ast }\right),\text{ \ \ }(\xi
_{1},\xi _{2})=a(\xi _{1}^{\ast },\xi _{2}^{\ast }),  \label{Eqn:1.nondim}
\end{equation}%
where $U_{0}=\sqrt{gh_{1}}$ is the characteristic speed of the gravity
waves. Here, all the variables with asterisks are assumed to be $O(1)$ in $%
\alpha $ and $\beta $. In the dimensionless, starred variables, Laplace's
equation (\ref{Eqn:1.Laplace}) is formulated as%
\begin{equation}
\beta \phi _{ix^{\ast }x^{\ast }}^{\ast }+\phi _{iz^{\ast }z^{\ast }}^{\ast
}=0,\ \ \ {\rm for\ } i=1,2.  \label{Eqn:1.non_Laplace}
\end{equation}

We first focus on the upper fluid layer. The analogous derivation for the
lower fluid layer follows a similar procedure. We seek an asymptotic
expansion of $\phi _{1}^{\ast }$ in powers of $\beta $,%
\begin{equation}
\phi _{1}^{\ast }=\phi _{1}^{\ast (0)}+\beta \phi _{1}^{\ast (1)}+O(\beta
^{2}),  \label{Eqn:1.expans}
\end{equation}%
which we use in the asymptotic analysis of the nondimensionalized problem of
equations (\ref{Eqn:1.Laplace})-(\ref{Eqn:1.dyn_if}) for small values of the
parameter $\beta $. From the leading $O(1)$ term in Eqs. (\ref%
{Eqn:1.Laplace}) and (\ref{Eqn:1.kin_sf}), $\phi _{1}^{\ast (0)}$ is found
to be independent of the height $z$,
\begin{equation}
\phi _{1}^{\ast (0)}=\phi _{1}^{\ast (0)}(x^{\ast },t^{\ast }).
\label{Eqn:1.phi1_zero}
\end{equation}%
The $O(\beta )$ terms in Eqs. (\ref{Eqn:1.Laplace}) and (\ref%
{Eqn:1.kin_sf}) yield the equations
\begin{equation*}
\phi _{1x^{\ast }x^{\ast }}^{\ast (0)}+\phi _{1z^{\ast }z^{\ast }}^{\ast
(1)}=0,\text{ \ \ }\alpha \xi _{2}^{\ast }<z^{\ast }<h_{1}^{\ast }+\alpha
\xi _{1}^{\ast },\text{ with }h_{1}^{\ast }=1,
\end{equation*}%
\begin{equation}
\phi _{1z^{\ast }}^{\ast (1)}=\xi _{1t^{\ast }}^{\ast }+\alpha \phi
_{1x^{\ast }}^{\ast (0)}\xi _{1x^{\ast }}^{\ast },\text{ \ \ at }z^{\ast
}=h_{1}^{\ast }+\alpha \xi _{1}^{\ast }.  \label{Eqn:1.phi1_one_eq}
\end{equation}%
The expression for $\phi _{1}^{\ast (1)}$ is obtained as

\begin{equation}
\phi _{1}^{\ast (1)}=D_{1}\xi _{1}^{\ast }Z_{1}-\frac{1}{2}\phi _{1x^{\ast
}x^{\ast }}^{\ast (0)}Z_{1}^{2},  \label{Eqn:1.phi1_one}
\end{equation}%
where $Z_{1}=z^{\ast }-h_{1}^{\ast }-\alpha \xi _{1}^{\ast }$ and $D_{1}\xi
_{1}^{\ast }\equiv \xi _{1t^{\ast }}^{\ast }+\alpha \phi _{1x^{\ast }}^{\ast
(0)}\xi _{1x^{\ast }}^{\ast }$. Combining expressions (\ref{Eqn:1.phi1_zero}%
) and (\ref{Eqn:1.phi1_one}), to the first power of $\beta $, we can obtain
the solution $\phi _{1}^{\ast }$ as
\begin{equation}
\phi _{1}^{\ast }=\phi _{1}^{\ast (0)}+\beta \left( D_{1}\xi _{1}^{\ast
}Z_{1}-\frac{1}{2}\phi _{1x^{\ast }x^{\ast }}^{\ast (0)}Z_{1}^{2}\right)
+O(\beta ^{2}).  \label{Eqn:1.phi1}
\end{equation}%
By integrating Eq. (\ref{Eqn:1.Laplace}) once from $\alpha \xi
_{2}^{\ast }$ to $h_{1}^{\ast }+\alpha \xi _{1}^{\ast }$ with respect to $%
z^{\ast }$, imposing the boundary conditions (\ref{Eqn:1.kin_sf}) and (%
\ref{Eqn:1.kin_if1}), and substituting the expression (\ref{Eqn:1.phi1})
into equation (\ref{Eqn:1.Laplace}), we obtain the kinematic equation for the upper fluid
layer,%
\begin{equation}
\eta _{1t^{\ast }}^{\ast }+\left( \eta _{1}^{\ast }\widetilde{u}_{1}^{\ast
}\right) _{x^{\ast }}-\frac{1}{6}\beta \left( h_{1}^{\ast }\right) ^{3}%
\widetilde{u}_{1x^{\ast }x^{\ast }x^{\ast }}^{\ast }-\frac{1}{2}\beta \left(
h_{1}^{\ast }\right) ^{2}\xi _{1t^{\ast }x^{\ast }x^{\ast }}+O(\alpha \beta
,\beta ^{2})=0,  \label{Eqn:1.kin_final}
\end{equation}%
where
\begin{equation*}
\eta _{1}^{\ast }=h_{1}^{\ast }+\alpha \xi _{1}^{\ast }-\alpha \xi
_{2}^{\ast },\text{ \ \ }\widetilde{u}_{1}^{\ast }=\phi _{1x^{\ast }}^{\ast
(0)}.
\end{equation*}%
Upon substitution of the velocity potential $\phi _{1}^{\ast }$ (\ref%
{Eqn:1.phi1}) into the dynamical boundary condition (\ref{Eqn:1.dyn_sf}), we
obtain the dynamical equation governing the motion of the upper fluid layer,%
\begin{equation}
\widetilde{u}_{1t^{\ast }}^{\ast }+\alpha \widetilde{u}_{1}^{\ast }%
\widetilde{u}_{1x^{\ast }}^{\ast }+g\xi _{1x^{\ast }}^{\ast }+O(\alpha \beta
,\beta ^{2})=0,  \label{Eqn:1.dyn_final}
\end{equation}%
where the equation has been differentiated with respect to $x^{\ast }$ once,
the terms in the first power of $\beta $ are retained [the $O(\beta)$ terms happen to
vanish in Eq. (\ref{Eqn:1.dyn_final})], and terms of $%
O(\alpha \beta ,\beta ^{2})$ are dropped. From the velocity potential $\phi
_{1}^{\ast }$, (\ref{Eqn:1.phi1}), we obtain the horizontal velocity $\phi
_{1x^{\ast }}^{\ast }$ as%
\begin{equation}
\phi _{1x^{\ast }}^{\ast }=\widetilde{u}_{1}^{\ast }+\beta \xi _{1t^{\ast
}x^{\ast }}^{\ast }Z_{1}-\frac{1}{2}\beta \widetilde{u}_{1x^{\ast }x^{\ast
}}^{\ast }Z_{1}^{2}+O(\alpha \beta ,\beta ^{2}).  \label{Eqn:1.phi1x}
\end{equation}%
By averaging (\ref{Eqn:1.phi1x}) over the depth, we obtain the layer-mean
horizontal velocity for the upper fluid layer,%
\begin{equation*}
\overline{u}_{1}^{\ast }=\widetilde{u}_{1}^{\ast }-\frac{1}{2}\beta
h_{1}^{\ast }\xi _{1t^{\ast }x^{\ast }}^{\ast }-\frac{1}{6}\beta \left(
h_{1}^{\ast }\right) ^{2}\widetilde{u}_{1x^{\ast }x^{\ast }}^{\ast
}+O(\alpha \beta ,\beta ^{2});
\end{equation*}%
the corresponding inverse is%
\begin{equation}
\widetilde{u}_{1}^{\ast }=\overline{u}_{1}^{\ast }+\frac{1}{2}\beta
h_{1}^{\ast }\xi _{1t^{\ast }x^{\ast }}^{\ast }+\frac{1}{6}\beta \left(
h_{1}^{\ast }\right) ^{2}\overline{u}_{1x^{\ast }x^{\ast }}^{\ast }+O(\alpha
\beta ,\beta ^{2}),  \label{Eqn:1.ubar}
\end{equation}%
where the layer-mean horizontal velocity is defined as
\begin{equation*}
\overline{u}%
_{1}^{\ast }(x^{\ast },t^{\ast })=\displaystyle\dfrac{1}{\eta _{1}^{\ast }}%
\int_{\alpha \xi _{2}^{\ast }}^{h_{1}^{\ast }+\alpha \xi _{1}^{\ast }}\phi
_{1x^{\ast }}^{\ast }(x^{\ast },z^{\ast },t^{\ast })dz^{\ast }.  \label{Eqn:uubar2}
\end{equation*}
After
substituting Eq. (\ref{Eqn:1.ubar}) for the horizontal velocity $\widetilde{u}_{1}^{\ast }$, equations (\ref{Eqn:1.kin_final}) and (\ref{Eqn:1.dyn_final}) provide Boussinesq-type equations governing the motion
of the fluid in the upper layer.

Repeating a similar procedure, we can obtain the governing equations for the
lower fluid layer. The final set of equations for the variables $(\xi _{1},\xi _{2},%
\overline{u}_{1},\overline{u}_{2})$, in the dimensional form, is%
\begin{equation}
\eta _{1t}+(\eta _{1}\overline{u}_{1})_{x}=0,\text{ \ \ }\eta _{1}=h_{1}+\xi
_{1}-\xi _{2},  \label{Eqn:1.upkin}
\end{equation}%
\begin{equation}
\eta _{2t}+(\eta _{2}\overline{u}_{2})_{x}=0,\text{ \ \ }\eta _{2}=h_{2}+\xi
_{2},  \label{Eqn:1.dnkin}
\end{equation}%
\begin{equation}
\overline{u}_{1t}+\overline{u}_{1}\overline{u}_{1x}+g\xi _{1x}-\frac{1}{3}%
h_{1}^{2}\overline{u}_{1xxt}+\frac{1}{2}h_{1}\xi _{2ttx}=0,
\label{Eqn:1.updyn}
\end{equation}%
\begin{align}
& \overline{u}_{2t}+\overline{u}_{2}\overline{u}_{2x}+g\xi _{2x}+\rho
_{r}g\eta _{1x}  \label{Eqn:1.dndyn} \\
& -\frac{1}{2}\rho _{r}h_{1}^{2}\overline{u}_{1xxt}+\rho _{r}h_{1}\xi
_{2ttx}-\frac{1}{3}h_{2}^{2}\overline{u}_{2xxt}=0,  \notag
\end{align}%
where $\rho _{r}$ is the density ratio $\rho _{1}/\rho _{2}$, and
\begin{equation*}
\overline{u}_{1}(x,t)=\dfrac{1}{\eta _{1}}\int_{\xi _{2}}^{h_{1}+\xi
_{1}}\phi _{1x}(x,z,t)dz, \ \ \ \overline{u}_{2}(x,t)=\dfrac{1}{\eta _{2}}\int_{-h_{2}}^{\xi _{2}}\phi
_{2x}(x,z,t)dz,
\end{equation*}%
are the layer-mean horizontal velocities.
We refer to the model (\ref{Eqn:1.upkin})-(\ref{Eqn:1.dndyn}) as the two-layer weakly-nonlinear model or TWN model.
The TWN model can also be obtained via a direct reduction
from a fully nonlinear model (to which we refer as MCC model) given in \cite{Choi1996JFM,Barros2007SAM,Barros2009SAM}.
For the numerical examples throughout the paper, all the variables and parameters are dimensionless and the parameters are fixed to be $(h_{1},h_{2},g,\protect\rho_{1},%
\protect\rho_{2}) = (1,3,1,1,1.003)$.
In particular, in our simulations, the characteristic length in both
height and wavelength is $h_{1}$, the characteristic speed is $\sqrt{gh_{1}}$%
, and the characteristic time is $\sqrt{h_{1}/g}$ [see similar dimensionless forms used in
numerical simulations in \cite{Choi1999JFM,Choi2009JFM}].
Note that, without loss of generality, our conclusions concerning broadening of IWs in Sec. \ref{subsec:traveling_wave} and asymmetric behavior of SWs in Sec. \ref{sec:surface_wave} also hold true for other parameter regimes of $(h_{1},h_{2},g,\protect\rho_{1},%
\protect\rho_{2})$.

\section{Basic properties of the TWN model}

\label{sec:basic_properties}

In this section, we study some basic properties of the TWN model, including interfacial solitary wave solutions and
dispersion relations.

\subsection{Interfacial solitary-wave solution}

\label{subsec:traveling_wave}

\begin{figure}
\centering  \hspace*{-2.0mm} %
\includegraphics[scale=0.65]{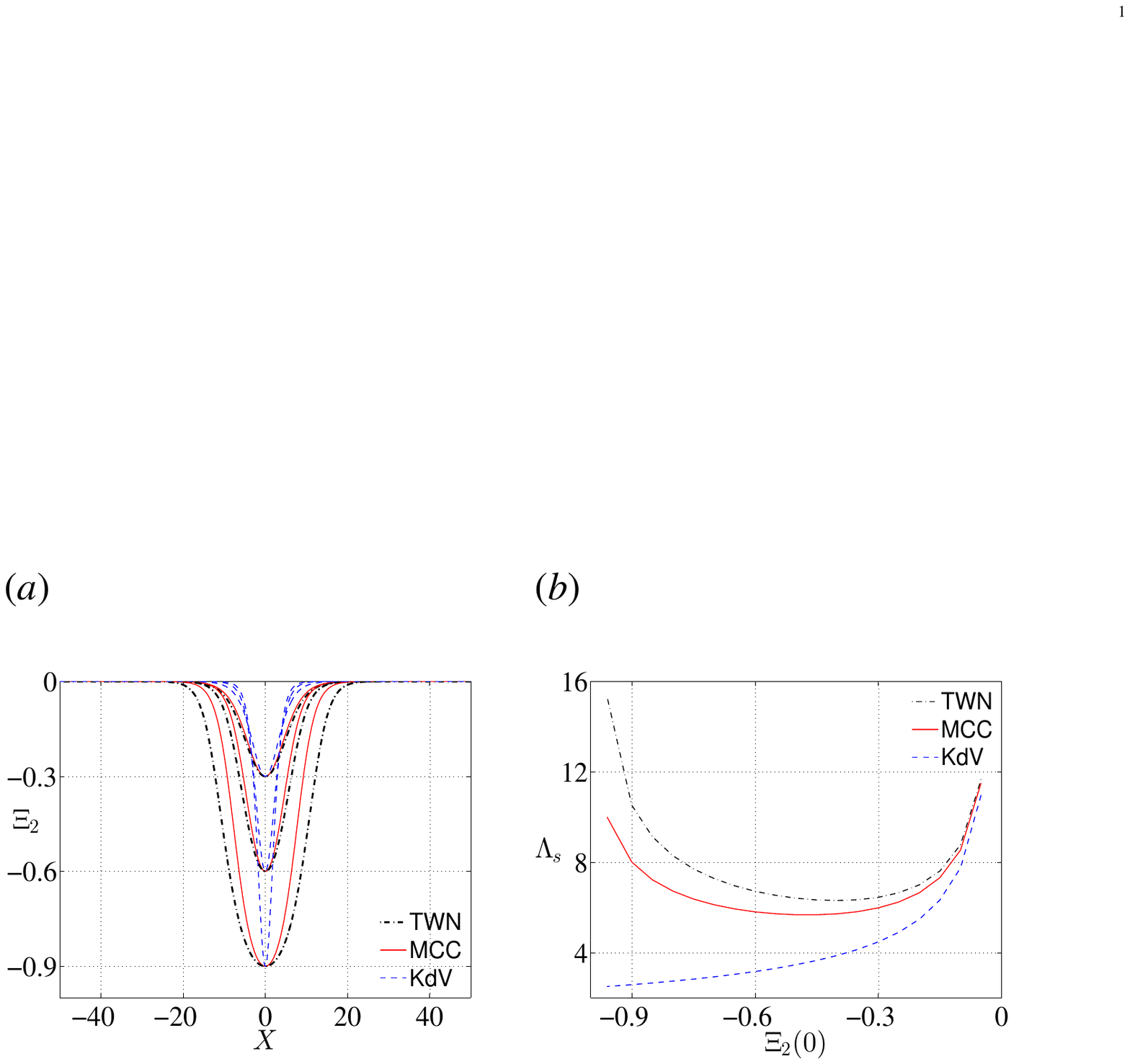}
\caption{(Color online) Comparison of interfacial solitary-wave
solutions among our TWN model [Eqs. (\protect\ref{Eqn:dynup})-(\protect\ref{Eqn:dyndn})],
the MCC model \cite{Choi1999JFM}, and the KdV model  \cite{Choi1999JFM}. (\textit{a}) Profiles of interfacial solitary-wave
solutions with the same amplitude.  (\textit{b})  Effective width $\Lambda_{s}$
versus wave amplitude $\protect\Xi_{2}(0)$. As the
amplitude increases, the solitary-wave solutions broaden and eventually
develop a flat crest when the amplitude increases to approximately $%
(h_{1}-h_{2})/2$. The TWN model and MCC model can capture the broadening of IWs whereas the KdV model
cannot.}
\label{Fig:broaden}
\end{figure}



%


To study the behavior of the overlaying SWs when an interfacial solitary wave moves
beneath the surface, we first seek interfacial solitary-wave solutions of
the TWN model (\ref{Eqn:1.upkin})-(\ref{Eqn:1.dndyn}). Then, we  use
these solitary-wave solutions as initial conditions for wave profiles and layer-mean velocities
for the subsequent numerical simulations in Sec. \ref{sec:surface_wave}.
To look for the right-moving traveling waves that propagate
with the nonlinear phase velocity $C_{s}$, we assume the following ansatz
for the surface elevation, internal elevation, upper-layer velocity, and
lower-layer velocity, $(\xi _{i},\overline{u}_{i})$ [$i=1,2$], in the system (\ref%
{Eqn:1.upkin})-(\ref{Eqn:1.dndyn}):
\begin{equation}
\xi _{i}\left( x,t\right) =\Xi _{i}\left( X\right) ,\text{ \ \ }\overline{u}%
_{i}\left( x,t\right) =\overline{U}_{i}\left( X\right) ,\text{ \ \ }%
X=x-C_{s}t.  \label{Eqn:ansatz}
\end{equation}%
Substituting this ansatz into Eqs. (\ref{Eqn:1.upkin})-(\ref%
{Eqn:1.dnkin}) and integrating once with respect to $X$ yields%
\begin{equation}
\overline{U}_{i}=\frac{C_{s}\left( H_{i}-h_{i}\right) }{H_{i}},\text{ with }%
H_{1}=h_{1}+\Xi _{1}-\Xi _{2}\text{ and }H_{2}=h_{2}+\Xi _{2},
\label{Eqn:Ui_bar}
\end{equation}%
where we have assumed that $H_{i}\rightarrow h_{i}$ as $X\rightarrow \pm
\infty $, and $h_{1}$ ($h_{2}$) is the undisturbed thickness of the upper
(lower) fluid layer, respectively. Substituting the horizontal velocity (\ref%
{Eqn:Ui_bar}) for $\overline{U}_{i}$ into Eqs. (\ref{Eqn:1.updyn})-(\ref%
{Eqn:1.dndyn}) and integrating once with respect to $X$ leads to the
equations%
\begin{align}
-\frac{1}{3}C_{s}^{2}h_{1}^{3}\frac{H_{1XX}}{H_{1}^{2}}-\frac{1}{2}%
C_{s}^{2}h_{1}H_{2XX}& \label{Eqn:dynup} \\
=\frac{C_{s}^{2}h_{1}^{2}}{2}\left( \frac{1}{H_{1}^{2}%
}-\frac{1}{h_{1}^{2}}\right) & +g\left( H_{1}+H_{2}-h_{1}-h_{2}\right)
 -\frac{2}{3}C_{s}^{2}h_{1}^{3}\frac{H_{1X}^{2}}{H_{1}^{3}},  \notag
\end{align}%
\begin{align}
-\frac{1}{2}\rho _{r}C_{s}^{2}h_{1}^{3}\frac{H_{1XX}}{H_{1}^{2}}-\rho
_{r}C_{s}^{2}h_{1}H_{2XX}-\frac{1}{3}C_{s}^{2}h_{2}^{3}\frac{H_{2XX}}{%
H_{2}^{2}}& \label{Eqn:dyndn} \\
=\frac{C_{s}^{2}h_{2}^{2}}{2}\left( \frac{1}{H_{2}^{2}}-\frac{1}{%
h_{2}^{2}}\right) +g\left( H_{2}-h_{2}\right)
+\rho _{r}g\left( H_{1}-h_{1}\right) & -\rho _{r}C_{s}^{2}h_{1}^{3}\frac{%
H_{1X}^{2}}{H_{1}^{3}}-\frac{2}{3}C_{s}^{2}h_{2}^{3}\frac{H_{2X}^{2}}{%
H_{2}^{3}},  \notag
\end{align}%
where we have assumed that $H_{iX}$, $H_{iXX}\rightarrow 0$ as $X\rightarrow
\pm \infty $ for $i=1,2$. Since explicit solutions to Eqs. (\ref%
{Eqn:dynup})-(\ref{Eqn:dyndn}) are difficult to establish, we numerically
compute their solitary-wave solutions by applying the method in \cite{Han2007AMC}.

In Fig. \ref{Fig:broaden}(\textit{a}), we
show the numerical solutions of the TWN model for IWs with different amplitudes. For
comparison, we also show the corresponding MCC  and KdV  solutions with the
same amplitudes \cite{Choi1999JFM}.
From Fig. \ref{Fig:broaden}(\textit{a}), we can see that the
TWN model and the MCC model can capture
the broadening of internal waves that is often observed in the ocean. For instance, a
single large internal wave in $340$ meters of water was observed in the northeastern South
China Sea by \cite{Duda2004IEEEJOE}. The typical wavelength of the observed internal wave is
longer than the KdV solution of the same amplitude that is used to fit this
internal wave. It is worthwhile to mention that the broadening of
interfacial solitary wave solutions can also be captured by other models %
\cite{Craig2005CPAM,Guyenne2006CRM}, not only by the MCC-type models.



To quantify this broadening, we introduce a measure of the
effective width, $\Lambda _{s}$, for the interfacial solitary-wave solution \cite{Koop1981JFM}, defined as
\begin{equation}
\Lambda _{s}=\left\vert \frac{1}{\Xi _{2}(0)}\int_{0}^{\infty }\Xi
_{2}(X)dX\right\vert .  \label{Eqn:soliton_speed}
\end{equation}%
Meanwhile, the effective width of MCC and KdV solutions are provided in the
reference \cite{Choi1999JFM}. Figure \ref{Fig:broaden}(\textit{b})
displays the comparison of effective width among
the TWN solutions, MCC solutions, and KdV solutions. When the amplitude of waves is small,
there is good agreement of the effective widths among all solutions.
However, when the amplitude of the waves becomes large, discrepancy grows rapidly among these
three solutions. When the amplitude increases
to the limiting value, approximately $%
(h_{1}-h_{2})/2$, the TWN and MCC solutions become much broader than the
KdV solutions.
The maximum amplitude of our TWN model is approximately $%
(h_{1}-h_{2})/2$. Beyond the maximum amplitude, no solitary waves can exist
for IWs.



\subsection{Dispersion relations }

\label{subsec:dispersion_absence_shear}

We now investigate the dispersion relation of the TWN model (\ref{Eqn:1.upkin})-(\ref{Eqn:1.dndyn}) using linear analysis. By substituting the
monochromatic solutions $(\xi _{i},\overline{u}_{i})\sim \exp [{\text i} (kx-\mu
_{k}t)]$ into the system (\ref{Eqn:1.upkin})-(\ref{Eqn:1.dndyn}), the pure linear
dispersion relation in the absence of shear between the frequency $\mu _{k}$
and the wavenumber $k$ can be obtained as
\begin{align}
& \left( 1+\rho _{r}k^{2}h_{1}h_{2}+\frac{1}{3}k^{2}h_{1}^{2}+\frac{1}{3}%
k^{2}h_{2}^{2}+\frac{1}{9}k^{4}h_{1}^{2}h_{2}^{2}+\frac{1}{12}\rho
_{r}k^{4}h_{1}^{3}h_{2}\right) \mu _{k}^{4}  \label{Eqn:dispersion_absence}
\\
& -\left( gh_{1}+gh_{2}+\frac{1}{3}gk^{2}h_{1}h_{2}^{2}+\frac{1}{3}%
gk^{2}h_{1}^{2}h_{2}\right) k^{2}\mu _{k}^{2}+\left( 1-\rho _{r}\right)
g^{2}k^{4}h_{1}h_{2}=0.  \notag
\end{align}%
Here, the shear is the interface and velocity jump induced by an interfacial solitary wave. The
same dispersion relation can be found in \cite{Barros2007SAM}.


\begin{figure}
\centering
\includegraphics[scale=0.6]{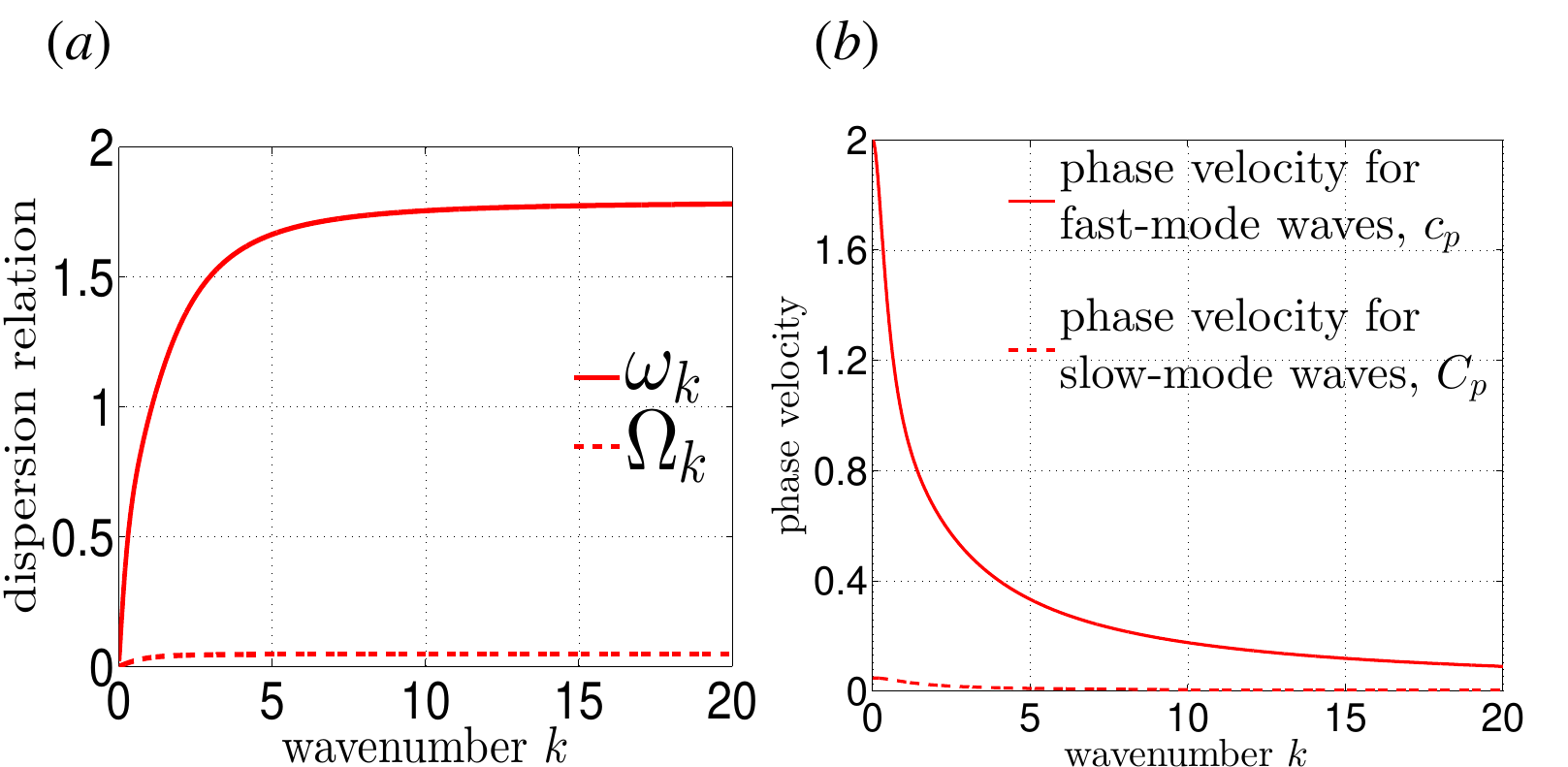}
\caption{(Color online) (\textit{a}) Pure linear dispersion relations $\Omega _{k}$ for slow-mode waves [Eq. (\ref{Eqn:disp_slow})] and $\omega _{k}$ for fast-mode waves
[Eq. (\ref{Eqn:disp_fast})]. (\textit{b}) Phase velocities $C _{p}$ for slow-mode waves and $c_{p}$ for fast-mode waves. }
\label{Fig:bare_dispersion2}
\end{figure}

Equation (\ref{Eqn:dispersion_absence}) has $4$ real roots  in the oceanic regime (the density
ratio $\rho _{r}$ is close to $1$). At the leading order in $1-\rho_{r}$,  the dispersion
relations of the two-mode waves, denoted by $\Omega _{k}$ and $\omega _{k}$ [see Fig. \ref{Fig:bare_dispersion2}],
can be approximated as
\begin{equation}
\Omega _{k}^{2}=\frac{\left( 1-\rho _{r}\right) g^{2}h_{1}h_{2}k^{2}}{%
gh_{1}+gh_{2}+\frac{1}{3}gk^{2}h_{1}h_{2}^{2}+\frac{1}{3}gk^{2}h_{1}^{2}h_{2}%
},  \label{Eqn:disp_slow}
\end{equation}%
and
\begin{equation}
\omega _{k}^{2}=\frac{\left( gh_{1}+gh_{2}+\frac{1}{3}gk^{2}h_{1}h_{2}^{2}+%
\frac{1}{3}gk^{2}h_{1}^{2}h_{2}\right) k^{2}}{1+\rho _{r}k^{2}h_{1}h_{2}+%
\frac{1}{3}k^{2}h_{1}^{2}+\frac{1}{3}k^{2}h_{2}^{2}+\frac{1}{9}%
k^{4}h_{1}^{2}h_{2}^{2}+\frac{1}{12}\rho _{r}k^{4}h_{1}^{3}h_{2}}.
\label{Eqn:disp_fast}
\end{equation}%
In the following, the two kinds of waves that correspond to the dispersion
relations $\Omega _{k}$, (\ref{Eqn:disp_slow}), and $\omega _{k}$,\ (\ref%
{Eqn:disp_fast}), are referred to as the slow-mode
waves and the fast-mode waves, respectively.





The modulated dispersion relation $\overline{\omega }_{k}$ in the presence of
shear can be obtained by substituting $(\xi _{i},\overline{u}%
_{i})\sim (\Xi _{i},\overline{U}_{i})+\exp [{\text i}(kx-\overline{\omega }_{k}t)]$
into the system (\ref{Eqn:1.upkin})-(\ref{Eqn:1.dndyn}), where the shear is induced by an interfacial solitary wave
$(\Xi _{i},\overline{U}_{i})$ in above Sec. \ref{subsec:traveling_wave}. The resulting equation is
\begin{equation}
\left\vert
\begin{array}{cccc}
-\overline{\omega }_{k}+k\overline{U}_{1} & \overline{\omega }_{k}-k%
\overline{U}_{1} & kh_{1}+k\Xi _{1}-k\Xi _{2} & 0 \\
0 & -\overline{\omega }_{k}+k\overline{U}_{2} & 0 & kh_{2}+k\Xi _{2} \\
gk & -\frac{1}{2}kh_{1}\overline{\omega }_{k}^{2} & -\overline{\omega }_{k}-%
\frac{1}{3}k^{2}h_{1}^{2}\overline{\omega }_{k}+k\overline{U}_{1} & 0 \\
\rho _{r}gk & (1-\rho _{r})gk-\rho _{r}kh_{1}\overline{\omega }_{k}^{2} & -%
\frac{1}{2}\rho _{r}k^{2}h_{1}^{2}\overline{\omega }_{k} & -\overline{\omega
}_{k}-\frac{1}{3}k^{2}h_{2}^{2}\overline{\omega }_{k}+k\overline{U}_{2}%
\end{array}%
\right\vert =0,  \label{modulated_DR}
\end{equation}%
where  $\left\vert \cdot
\right\vert $ denotes the determinant of the enclosed matrix.
In the following Secs. \ref{sec:scheme} and \ref{sec:surface_wave}, we will
numerically study the TWN model in the right-moving frame with the nonlinear phase
velocity $C_{s}$, that is, $T=t$ and $X=x-C_{s}t$. Note that the solitary-wave solutions $(\Xi _{i},\overline{U}_{i})$ are
steady in time $T$ in this moving frame.
Then, the modulated dispersion
relation $\overline{\nu }$, corresponding to  the moving frame $T$ and $X$,  is given by
\begin{equation}
\overline{\nu }_{k}=\overline{\omega }_{k}-C_{s}k, \label{Eqn:nu_omega}
\end{equation}
where
$\overline{\omega }_{k}$ is the modulated dispersion relation in Eq. (\ref{modulated_DR}) corresponding to the resting frame $t$ and $x$.
Note that the dispersion relation $\overline{\nu }_{k}$ is independent of time $T$ since $(\Xi _{i},\overline{U}_{i})$ are steady in time $T$.
Moreover, the
wavelengths of $(\Xi _{i},\overline{U}_{i})$ are relatively
long with respect to the characteristic wavelengths of fast-mode waves, so $(\Xi _{i},\overline{U}%
_{i})$ in the dispersion relation $\overline{\nu }_{k}$ can be
locally treated as constant in $X$ space.

\section{Numerical scheme}

\label{sec:scheme}

\begin{figure}
\centering 
\includegraphics[scale=0.58]{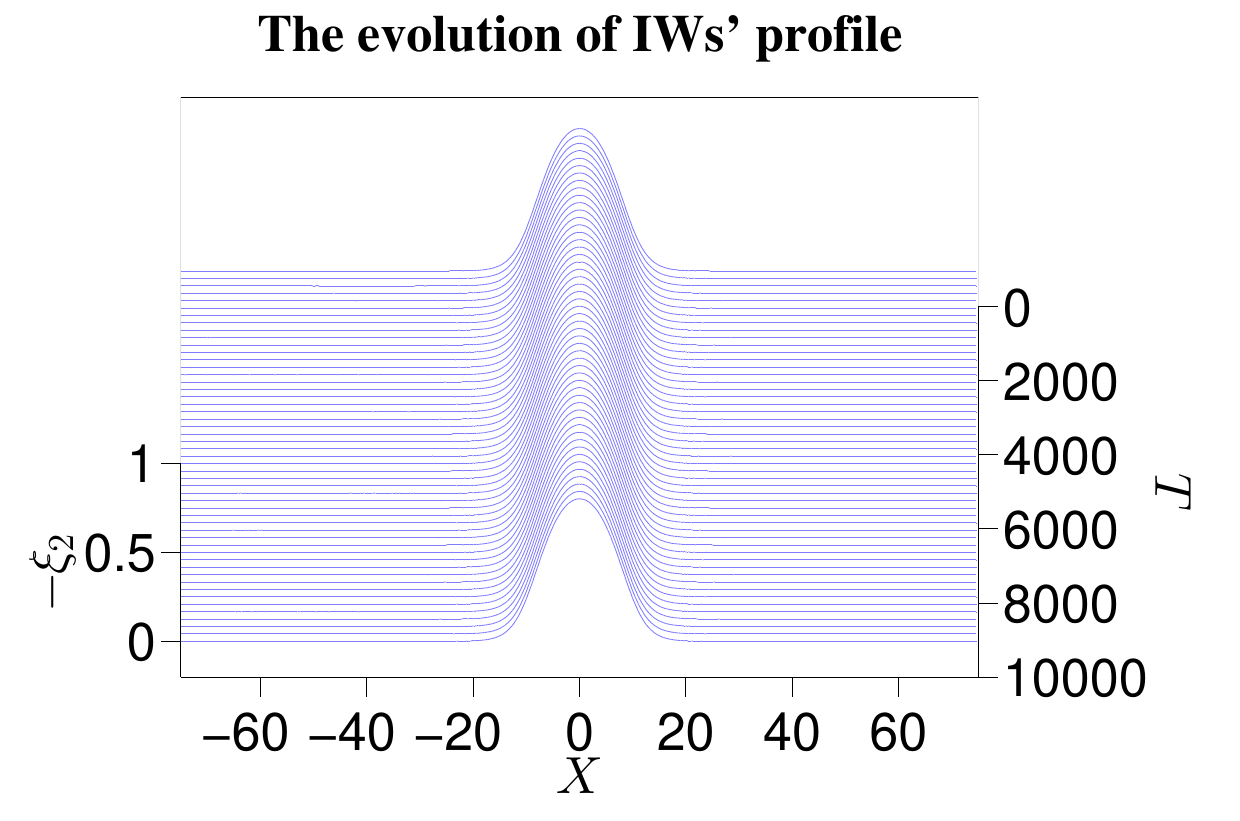}
\caption{(Color online) Spatiotemporal evolution of the interfacial solitary-wave solution $\protect\xi_{2}$ with the amplitude being $-0.8$ during the time period $0 \leq T \leq10000$. The maximal amplitudes of  $(\xi _{1},\xi _{2},\overline{u}_{1},\overline{u}_{2})$ are $(0.001,-0.8,0.024,-0.019)$. The
time step is $\Delta T = 0.1$ and the spatial discretization is $\Delta X = 600/2^{14}$.}
\label{Fig:Internal_Evol}
\end{figure}

For numerical computations, we cast Eqs. (\ref{Eqn:1.upkin})-(\ref%
{Eqn:1.dndyn}) in the conservation form in the right-moving frame with the
nonlinear phase velocity (soliton speed) $C_{s}$ as follows:%
\begin{equation}
\eta _{1T}-C_{s}\eta _{1X}+(\eta _{1}\overline{u}_{1})_{X}=0,
\label{Eqn:eta1_moving}
\end{equation}%
\begin{equation}
\eta _{2T}-C_{s}\eta _{2X}+(\eta _{2}\overline{u}_{2})_{X}=0,
\label{Eqn:eta2_moving}
\end{equation}

\begin{equation}
M_{1T}-C_{s}M_{1X}+\left( \frac{1}{2}\overline{u}_{1}^{2}+g\xi _{1}\right)
_{X}=0,  \label{Eqn:M1_moving}
\end{equation}%
\begin{equation}
M_{2T}-C_{s}M_{2X}+\left( \frac{1}{2}\overline{u}_{2}^{2}+g\xi _{2}+\rho
_{r}g\eta _{1}\right) _{X}=0,  \label{Eqn:M2_moving}
\end{equation}%
where
\begin{equation}
M_{1}=\overline{u}_{1}-\frac{1}{3}h_{1}^{2}\overline{u}_{1XX}-\frac{1}{2}%
h_{1}\left( \eta _{2}\overline{u}_{2}\right) _{XX},  \label{Eqn:M1_def}
\end{equation}%
\begin{equation}
M_{2}=\overline{u}_{2}-\frac{1}{2}\rho _{r}h_{1}^{2}\overline{u}_{1XX}-\frac{%
1}{3}h_{2}^{2}\overline{u}_{2XX}-\rho _{r}h_{1}\left( \eta _{2}\overline{u}%
_{2}\right) _{XX},  \label{Eqn:M2_def}
\end{equation}%
and $T=t$, $X=x-C_{s}t$.
The computational domain is set to be $[-M, M]$, with periodic boundary conditions.
Even for an initially narrowly localized perturbation
wave, radiation can be quickly emitted towards the two
boundaries $x=-M$ and $x=M$. To eliminate possible reflected waves from
these boundaries,  two buffer zones in the regions $[-M,-M/2]$
and $[M/2,M]$ are established, and  damping and diffusion terms are added to absorb the outgoing
radiation. For numerical integration, we use the fourth-order Runge-Kutta
method in time and the second-order collocation method in space \cite{Chen2005Efficient}.
The Kelvin-Helmholtz (KH) instability is suppressed by applying a low-pass filter %
\cite{Jo2008SAM}. (The wavenumbers for the KH instability are much larger than the wavenumbers of the trapped right-moving SWs as introduced in the following Sec. \ref{sec:surface_wave}. Thus, these KH unstable wavenumbers are physically irrelevant in our computations.) In our simulations, we fix the parameter regime $(\rho _{1},\rho
_{2},h_{1},h_{2},g)=(1,1.003,1,3,1)$ and the computational domain $M=300$.
All the variables and parameters in our simulations are
dimensionless.

We first focus on the evolution of initially unperturbed interfacial solitary-wave solutions. Figure \ref{Fig:Internal_Evol}
shows the spatiotemporal evolution of the IWs' profile, $\xi _{2}$, for $0\leq T\leq 10000$. We can see from Fig. \ref{Fig:Internal_Evol} that the IWs maintain their shape while traveling. This
result is consistent with many experimental observations that
large-amplitude internal waves typically can propagate over long distances with their
shape virtually unchanged \cite{Helfrich2006ARFM}.




\begin{figure}
\centering 
\includegraphics[scale=0.68]{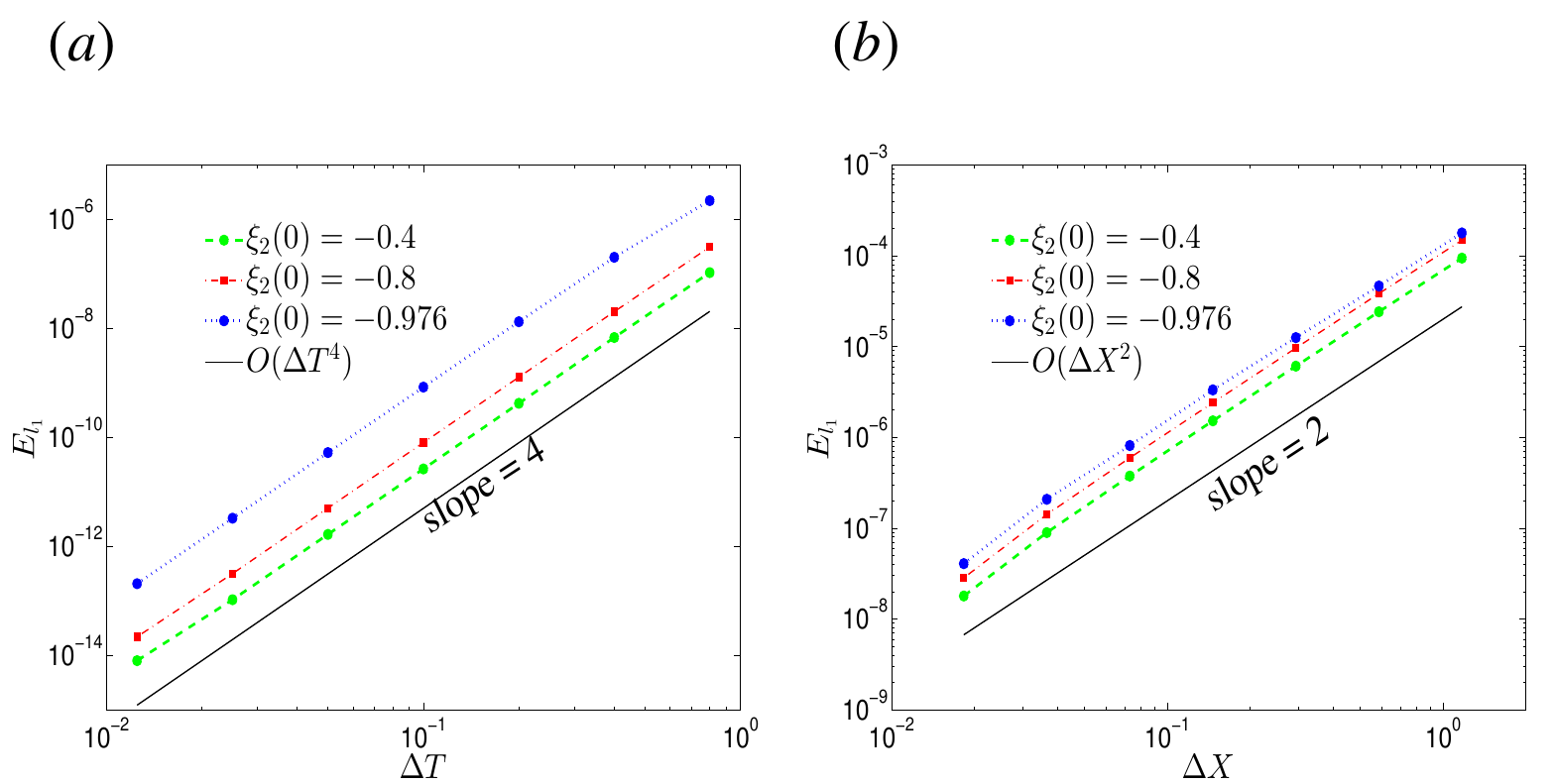}
\caption{(Color online) Numerical convergence examination of the scheme
in time and space. (\textit{a}) The $l_{1}$ norm error (\ref{l1_norm}) as a function
of the time step $\Delta T$ under different-amplitude IWs. The green, red, and blue lines correspond
to the IW amplitudes $\protect\xi%
_{2}(0)$ being $-0.4$, $-0.8$, and $-0.976$, respectively. To examine the convergence in time, we use a stopping time $T_{s}=4$ and fix
a spatial discretization $\Delta X = 600/2^{10}$. The reference solution is computed with a very small
time step $\Delta T = 0.1/32$.  The result in panel (\textit{a}) shows fourth-order time-accuracy
of the scheme. (\textit{b}) The $l_{1}$ norm error (\ref{l1_norm}) as a
function of the spatial discretization $\Delta X$ under different-amplitude IWs.
To examine the convergence in space, we use a stopping time $T_{s}=1$
and fix the time step $\Delta T = 0.01$.
The reference solution is computed with $\Delta X =
600/2^{16}$.
The result in panel (\textit{b}) shows second-order accuracy in
space of the scheme. }
\label{Fig:Accuracy_Test}
\end{figure}


Next, we examine the numerical convergence in time and space of our scheme for initially unperturbed
interfacial solitary-wave solutions [Eqs. (\protect\ref{Eqn:dynup})-(\protect\ref{Eqn:dyndn})]. We compute the $l_{1}$
norm error for the SWs' profile defined as%
\begin{equation}
E_{l_{1}}=\sum_{j}\left\vert \xi _{1}(X_{j})-\xi
_{1}^{\rm ref}(X_{j})\right\vert \Delta X,  \label{l1_norm}
\end{equation}%
where the reference solution $\xi _{1}^{\rm ref}$ is approximated by the numerical result
obtained from a very
small time step for the time accuracy test or from a very small spatial discretization for the spatial accuracy test.
We can see from Fig. \ref{Fig:Accuracy_Test} that the scheme has fourth-order accuracy in
time and second-order accuracy in space.


\section{Asymmetric behavior of SWs in the presence of an underlying IW}

\label{sec:surface_wave}




\begin{figure}
\centering 
\includegraphics[scale=0.275]{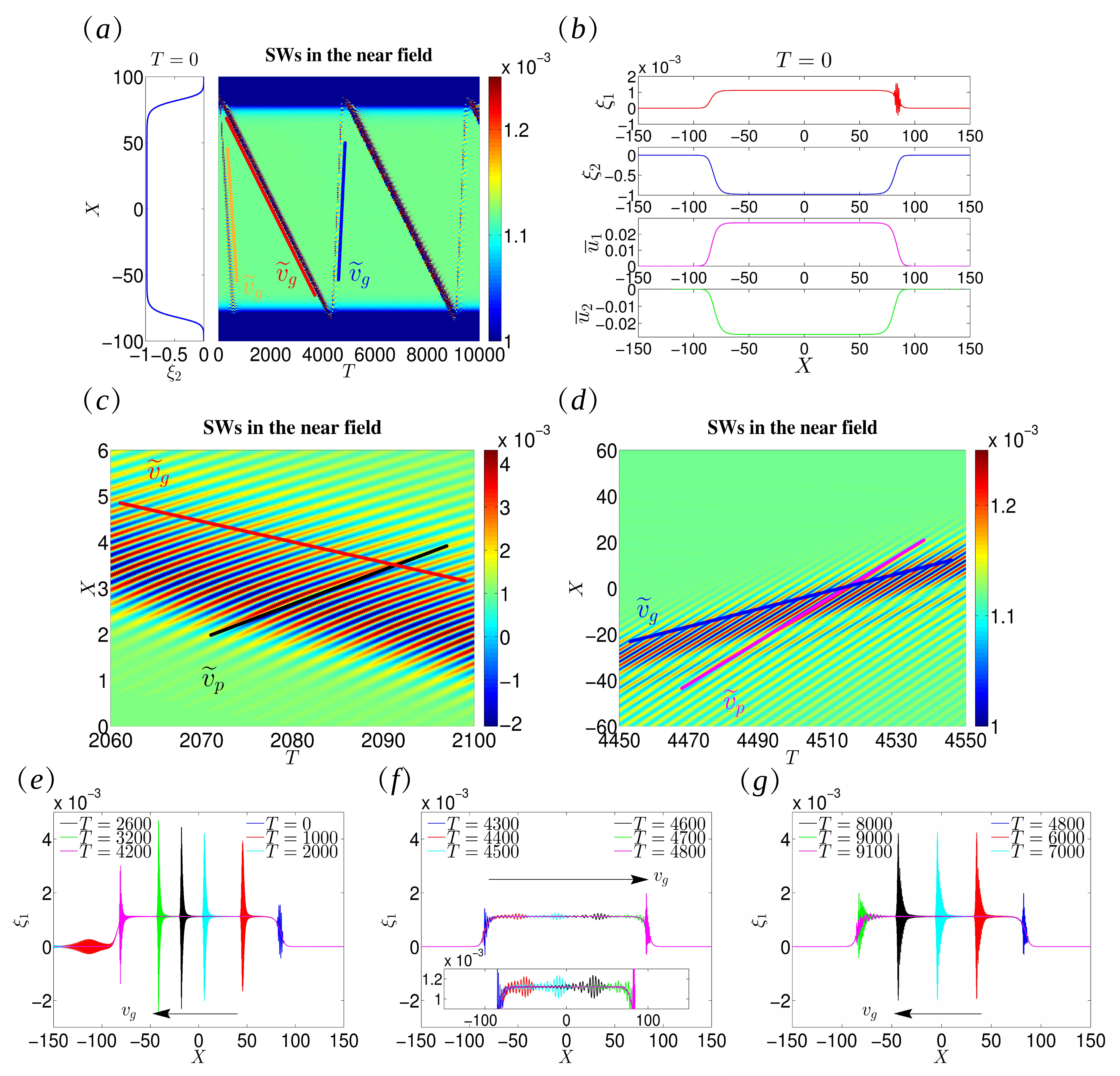}
\caption{(Color online) (\textit{a}) Spatiotemporal evolution of SWs'
profile $\protect\xi _{1}$ in the near field for $0\leq T\leq 10000$ and $%
-100\leq X\leq 100$, and snapshot of the interfacial solitary wave at $T=0$. The amplitude of the interfacial solitary wave is $-0.976$.  The yellow line corresponds to the group velocity of the left-moving fast-mode SW packet that is not trapped
in the near field. The red (blue) line corresponds to the
negative (postive) group velocity, $\widetilde{v}_{g}$, when the right-moving fast-mode SW packet propagates towards the trailing (leading) edge. (\textit{b}) The initial condition for wave profiles and horizontal velocities $%
(\protect\xi _{1},\protect\xi _{2},\overline{u}_{1},\overline{u}_{2})$. (%
\textit{c}) Zoomed-in version of panel (\textit{a}) for $2060\leq T\leq 2100$ and $%
0\leq X\leq 6$.
The black line corresponds to the positive phase
velocity, $\widetilde{v}_{p}$, and the red line corresponds to the negative group velocity, $\widetilde{v}_{g}$.
Wave packets are traveling in the direction of decreasing $X$ towards the trailing edge.
(\textit{d}) Zoomed-in version of panel (\textit{a}) for $%
4450\leq T\leq 4550$ and $-60\leq X\leq 60$. The magenta line corresponds to the positive phase
velocity, $\widetilde{v}_{p}$, and the blue line corresponds to the positive group velocity, $\widetilde{v}_{g}$.
Wave packets are traveling in
the direction of increasing $X$ towards the leading edge. (\textit{e})(\textit{f}%
)(\textit{g}) Snapshot of SWs in the near field.
In panels (\textit{e}) and
(\textit{g}), the group velocities $v_{g}$ are negative. In
panel (\textit{f}),  the group velocity $v_{g}$ is positive.}
\label{Fig:asymmetric}
\end{figure}






In this section, we present our numerical results for the system (\ref%
{Eqn:1.upkin})-(\ref{Eqn:1.dndyn}) describing the  behavior of a SW packet in the presence of an underlying interfacial solitary wave, and then compare them to the results of our theoretical
analysis using ray-based theories.
First, we initialize the SWs' height $\xi _{1}$ to be a profile
composed of a sufficiently-long-wavelength interfacial-solitary-wave solution and a localized perturbation,
that is, the initial condition [Fig. \ref{Fig:asymmetric}(\textit{b})]
for $(\xi _{1},\xi _{2},\overline{u}_{1},\overline{u}_{2})$ is taken to be
\begin{equation}
(\Xi _{1}+\delta _{1},\Xi _{2},\overline{U}_{1},\overline{U}_{2}),
\label{Eqn:initial_broad}
\end{equation}%
where
$\left(
\Xi _{1},\Xi _{2},\overline{U}_{1},\overline{U}_{2}\right) $ is the
solitary-wave solutions described in \S \S\ \ref{subsec:traveling_wave}
and
the localized perturbation $\delta_{1}$ is a narrow SW packet with a narrow band of
wavenumbers,
\begin{equation}
\delta _{1}=A_{\varepsilon }[\tanh (X-X_{0}+x_{0})-\tanh
(X-X_{0}-x_{0})]\cos (k_{0}(X-X_{0})),  \label{Eqn:perturb}
\end{equation}
with $A_{\varepsilon }=5\times 10^{-4}$, $X_{0}=84$, $x_{0}=2$, and $k_{0}=5$. For the interfacial solitary wave, $\Xi _{2}$, the
amplitude is $-0.976$, the
wavelength is $\sim 150$, and the speed is $C_{s}=0.0549$.

Below, the group velocity $v_{g}$, the phase velocity $v_{p}$, and the
frequency $\nu $ correspond to the moving frame $(T,X)$.
On the other hand, the group velocity $c_{g}=(v_{g}+C_{s})$, the phase velocity $c_{p}=(v_{p}+C_{s})$%
, and the frequency $\omega (=\nu+C_{s}k)$ correspond to the resting frame $(t,x)$.
The variable with the tilde, $\widetilde{\cdot}$, stands for the numerical
measurement of the corresponding quantity.


Initially, left-moving radiation is quickly emitted from the
near field and eventually absorbed by our absorbing boundary condition in the buffer zones [dark stripe parallel to the yellow line in Fig. \ref{Fig:asymmetric}(\textit{a})].
After this initial
transient, we can see that one SW packet is
trapped in the near field [dark stripes parallel to the red line and blue line in Fig. \ref{Fig:asymmetric}(\textit{a})].
These trapped waves are all
right-moving waves, that is, their phase velocities $%
\widetilde{v}_{p}>0$ are positive [dark stripes parallel to the black line in Fig. \ref{Fig:asymmetric}(\textit{c})
and to the magenta line in Fig. \ref{Fig:asymmetric}(\textit{d})].
Thus, only the right-moving SWs that propagate in the
same direction as the underlying interfacial solitary wave remain trapped in the near field.

We now study the spatiotemporal manifestation of these right-moving SWs in the near field.
From Fig. \ref{Fig:asymmetric}, we can observe three features of these right-moving SWs:

(i) SWs become short in wavelength at the leading edge and long at the
trailing edge. For $0\leq T\leq 4300$, the SW packets propagate towards the trailing edge with a relatively large wavenumber
$\widetilde{k}\sim 23$ [Fig. \ref{Fig:asymmetric}(\textit{e})].
For $4300\leq T\leq 4700$, the
SW packets propagate towards the leading edge with a relatively small wavenumber $%
\widetilde{k}\sim 1.8$ [Fig. \ref{Fig:asymmetric}(\textit{f})].

(ii) SW packets propagate towards the trailing edge with a relatively small
group velocity, and towards the leading edge with a relatively large group velocity.
From Figs. \ref{Fig:asymmetric}(\textit{a})(\textit{c})(\textit{e}), we can
see that for $0\leq T\leq 4300$, the SW packets
propagate towards the trailing edge with a relatively small group
velocity  $\widetilde{v}_{g}\sim -0.045$.
For $4300\leq T\leq 4700$ [Figs. \ref{Fig:asymmetric}(\textit{a})(\textit{d})(\textit{f})], the SW packets propagate towards the leading edge
with a relatively large group velocity $\widetilde{v}_{g}\sim 0.37$.
For $4800\leq T\leq 9100$ [Figs. \ref{Fig:asymmetric}(\textit{a})(\textit{g})], the SW packets again propagate towards the trailing edge with a relatively small group velocity.

(iii) SWs become high in amplitude at the leading edge and low at the trailing
edge. For $0\leq T\leq 4300$, the SW packets' amplitude increases to $\sim
3\times 10^{-3}$ at the leading edge and then SWs propagate towards the trailing edge
[Fig. \ref{Fig:asymmetric}(\textit{e})]. For $4300\leq T\leq 4700$, the
SW packets' amplitude decreases to $\sim 1\times 10^{-4}$ at the trailing edge
and then propagate  towards the leading edge
[Fig. \ref%
{Fig:asymmetric}(\textit{f})].

\begin{figure}
\center \hspace*{0mm} 
\includegraphics[height=3 in]{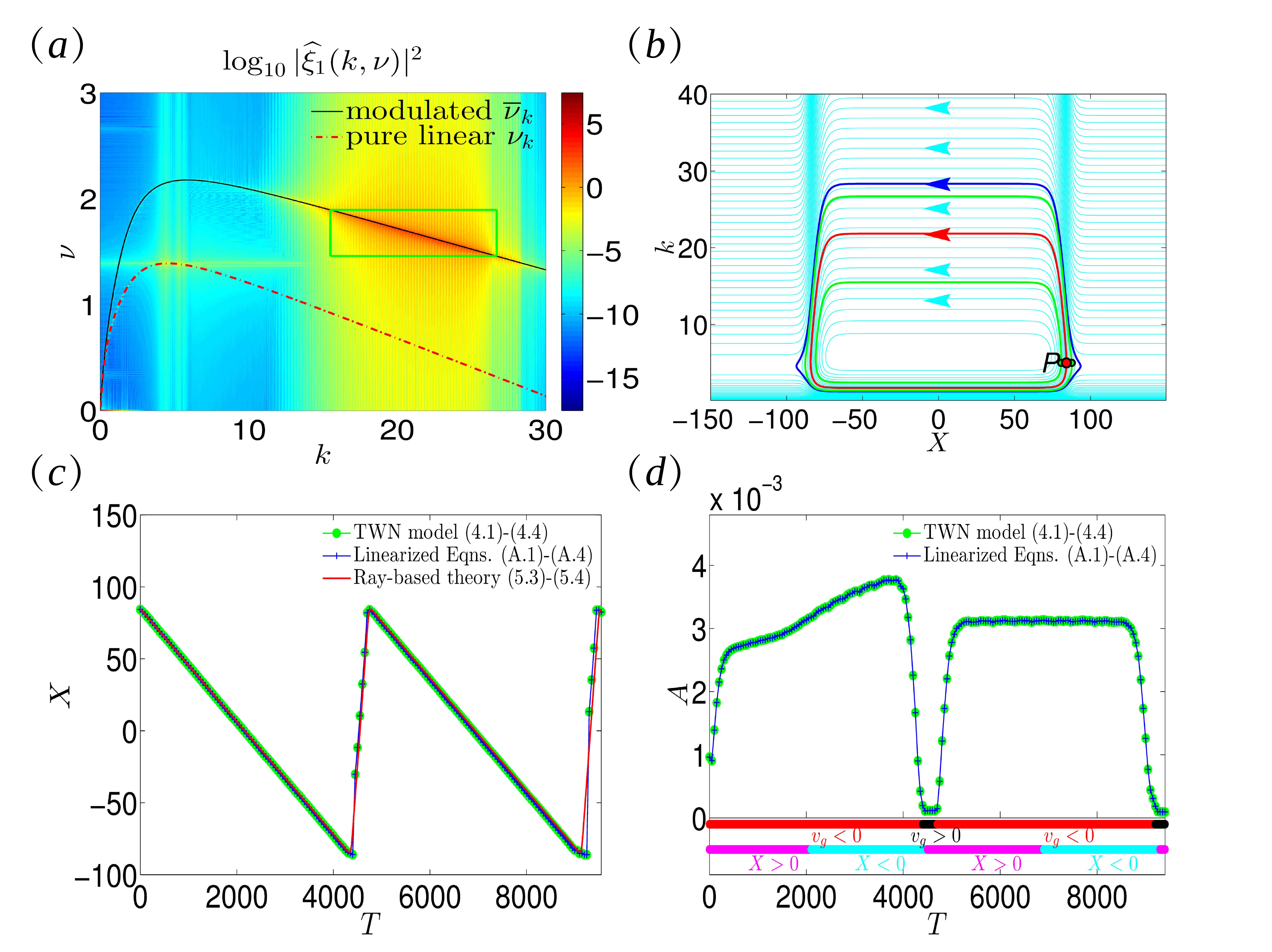}
\caption{(Color online) (\textit{a}) The logarithmic modulus, ${\ \log_{10}
|\widehat{\protect\xi }_{1}(k,\protect\nu )|^{2}}$, of SWs' profile $\protect%
\xi _{1}$ for $3181\leq T\leq 4000$ and $-150\leq X\leq 150$. For
comparison, also plotted are the pure linear dispersion relation $\protect\nu _{k} $  (red dashed-dotted curve) and the modulated dispersion relation $\overline{\protect\nu }%
_{k}$  (black solid curve). The green rectangle
corresponds to the range of wavenumbers and frequencies predicted by the
ray-based theories [see text and panel (\textit{b}) for details]. (\textit{b})
The phase portrait of the motion of wave packets in variables $X$ and $k$.
Along each cyan-colored curve, the frequency $\overline{\protect\nu }_{k}$
remains constant. Inside the region enclosed by the blue-level curve, the
fast-mode waves are trapped. The red point $P$, $(X,k)=(84,5)$, corresponds to
the central location and dominant wavenumber of
the perturbation (\protect\ref{Eqn:perturb}) and the red-level curve corresponds to the wave passing
through the point $P$ with a constant frequency. The distance of two
neighboring green points corresponds to the spatial width of the initial
perturbation (\protect\ref{Eqn:perturb}) and the two green-level curves
correspond to the wave passing through the two green points with constant
frequencies $\protect\nu = 1.46$ and $\protect\nu=1.895$.
On these two green-level curves, the maximal wavenumbers are $26.7$ and $15.5$, respectively.
For the green rectangle in panel (\textit{a}), the upper and lower bounds
correspond to the two frequencies $\protect\nu = 1.46$ and $\protect\nu%
=1.895 $ and the left and right bounds correspond to the wavenumbers $k=26.7$
and $k=15.5$. The arrows indicate the direction of movement of wave
packets. (\textit{c}) The temporal evolution of peak location of the fast-mode
waves in the near field. The green curve corresponds to the peak locations of
the fast-mode waves from the TWN model (\protect\ref{Eqn:eta1_moving})-(\protect
\ref{Eqn:M2_moving}). The blue curve corresponds to the peak locations of the
fast-mode waves from the effective linearized equations (\protect\ref%
{Eqn:eff_1})-(\protect\ref{Eqn:eff_4})
in appendix \ref{effective_linear_eqn}. The red curve corresponds to the
peak locations of the wave packet predicted by the ray-based theory (\protect\ref%
{Eqn:ray_base})-(\protect\ref{Eqn:wavenum_conserve}). They
nearly overlap one another. (\textit{d}) The
temporal evolution of the maximal amplitude, $A$, of the fast-mode waves. The
green curve and the blue curve correspond to the height of $A$ for the
fast-mode waves of the TWN model (\protect\ref{Eqn:eta1_moving})-(\protect\ref%
{Eqn:M2_moving}) and those of the effective
linearized equations (\protect\ref{Eqn:eff_1})-(\protect\ref{Eqn:eff_4})
in appendix \ref{effective_linear_eqn},
respectively. They overlap one another. The time period
marked by the red line corresponds to the negative group velocity $v_{g}$
and the black line corresponds to the positive group velocity $v_{g}$.
The time period marked by the magenta line corresponds to the positive peak
location $X$ and the cyan-line one corresponds to the negative peak location
$X$.}
\label{Fig:Broad_near_field_theory}
\end{figure}



To understand the dynamical
behavior of these right-moving SWs, we first quantify the dispersion relation of these waves.
Figure \ref{Fig:Broad_near_field_theory}(\textit{a}) shows the logarithmic modulus, $%
\log _{10}|\widehat{\xi }_{1}(k,\nu )|^{2}$, with its magnitude color-coded,
where $\widehat{\xi }_{1}(k,\nu )$ is the spatiotemporal Fourier transform
of $\xi _{1}(X,T)$. For comparison, also shown are the pure linear dispersion relation $%
\nu _{k}=\omega _{k}-C_{s}k$  [Eq. (\ref{Eqn:disp_fast}), red dashed-dotted
curve in Fig. \ref{Fig:Broad_near_field_theory}(\textit{a})]
and the modulated dispersion relation $\overline{\nu }_{k}=\overline{%
\omega }_{k}-C_{s}k$ [Eq. (\ref{modulated_DR}), black solid curve in Fig. \ref{Fig:Broad_near_field_theory}(\textit{a})].
For the modulated dispersion relation $\overline{\nu }_{k}$, we take
the amplitude of the interfacial solitary wave  to be $(\Xi _{1},\Xi _{2},\overline{U}_{1},\overline{U%
}_{2})=(0.001,-0.976,0.027,-0.026)$.
We can clearly see from
Fig. \ref{Fig:Broad_near_field_theory}(\textit{a}) that, for $15\leq k\leq
25$, the modulated dispersion relation $\overline{\nu }_{k}$ can
capture the peak locations of the spectrum well, whereas the pure linear dispersion relation $%
\nu _{k}$ deviates greatly.
Therefore, these right-moving SWs
can be well characterized by the modulated dispersion relation $\overline{%
\nu }_{k}$ and thereafter referred to as right-moving fast-mode SWs
[see the definition of fast-mode waves in Sec. \ref{subsec:dispersion_absence_shear}].

Incidentally, there are two yellow spots on the pure linear dispersion
relation $\nu _{k}$ near $k=5$, as can be observed faintly in Fig.
\ref{Fig:Broad_near_field_theory}(\textit{a}). However,
the spectral power at these two yellow spots on the linear dispersion
relation $\nu _{k}$ is six orders of
magnitude lower than that at
the modulated dispersion relation $\overline{\nu }_{k}$.
These two blurry yellow spots correspond to the wave
spectra of radiation waves in the far field, which is not the interest of this work.







To understand the asymmetric behavior (i)-(iii),  we compare our
numerical results of the TWN model
with those of the effective linearized equations (\ref{Eqn:eff_1})-(\ref{Eqn:eff_4}) in appendix \ref{effective_linear_eqn}.
Effective linearized equations (\ref{Eqn:eff_1})-(\ref{Eqn:eff_4}) are obtained from the linearization of our TWN model (\ref{Eqn:eta1_moving})-(\ref{Eqn:M2_moving}) in the presence of the interfacial solitary wave. Their
mathematical details are presented in appendix \ref{effective_linear_eqn}.
We also compare our TWN solutions with the theoretical predictions of ray-based theories. Due
to the slow varying in space and time of the phase of fast-mode waves, the
governing equations of space-time rays for the location $X$ and the
wavenumber $k$ are given by \cite{Whitham1974Linear,Bakhanov2002JGR},%
\begin{align}
\frac{dX}{dT}&=\frac{\partial \overline{\nu }_{k}}{\partial k},
\label{Eqn:ray_base}\\
\frac{dk}{dT}&=-\frac{\partial \overline{\nu }_{k}}{\partial X},
\label{Eqn:wavenum_conserve}
\end{align}%
where $\overline{\nu }_{k}$ is the modulated dispersion relation  in
Eq. (\ref{modulated_DR}).
Note that the modulated dispersion relation $\overline{\nu }_{k}$ does not
explicitly depend on time $T$ since the interfacial solitary wave  is
stationary in the moving frame. Clearly equations (\ref{Eqn:ray_base}) and (\ref%
{Eqn:wavenum_conserve}) constitute a Hamiltonian system with $\overline{\nu }%
_{k}$ as the Hamiltonian, $X$ displacement and $k$ momentum.
Equation (\ref{Eqn:ray_base}) states that the
wave packet propagates at the group velocity.




We now quantify the asymmetric behavior of these right-moving fast-mode SWs
by comparing the results of the TWN model, the results of the effective linearized equations,
and theoretical predictions from ray-based theories:

(i) First, we provide a theoretical prediction for the temporal evolution of the
wavenumber (equivalently the wavelength) of fast-mode waves. By the ray-based theory,
when propagating with the initial perturbation (\protect\ref{Eqn:perturb}), fast-mode waves
possess the peak locations $X$ and wavenumbers $k$ between the two green-level curves in Fig. \ref{Fig:Broad_near_field_theory}(\textit{b}).
For the minimal
wavenumber, the theoretical prediction $k=1.7$ can be attained at $X=0$
on the red-level curve  in Fig. \ref{Fig:Broad_near_field_theory}(\textit{b}).
This theoretical minimal
wavenumber is in good agreement with the measured one
$\widetilde{k}\sim 1.8$ at $(T,X)=(4500,0)$ in Fig. \ref{Fig:asymmetric}(\textit{f}).
For the maximal wavenumber, the theoretical prediction ranges from $15.5$ to $26.7$ between
the two green-level curves in Fig. \ref{Fig:Broad_near_field_theory}(\textit{b}).
For the theoretical wavenumbers $k=15.5$ and $k=26.7$, the corresponding frequencies
are $\protect\nu=1.895$ and $\protect\nu = 1.46$, respectively.
This range of theoretical wavenumbers and frequencies [depicted by the green
rectangle in Fig. \ref{Fig:Broad_near_field_theory}(\textit{a})] is in good agreement
with the the range of the measured ones in the spectrum in Fig. \ref{Fig:Broad_near_field_theory}(\textit{a}).
Therefore, the temporal evolution of the wavenumber can be characterized by the
ray-based theory (\ref{Eqn:ray_base})-(\ref{Eqn:wavenum_conserve}) for
fast-mode waves. In particular, fast-mode SWs
become short in wavelength at the leading edge [$X>0$] and long at the
trailing edge [$X<0$] [see Fig. \ref{Fig:Broad_near_field_theory}(\textit{b})].



(ii) Next, we investigate the motion of the peak location $X$ as a function of time $T$.
Figure \ref{Fig:Broad_near_field_theory}(\textit{c}) displays the temporal evolution of
numerically measured peak locations of the fast-mode waves for the TWN model (\ref{Eqn:eta1_moving})-(\ref{Eqn:M2_moving}) [green curve in Fig. \ref{Fig:Broad_near_field_theory}(\textit{c})] as well as the prediction using the effective linearized equations (\ref{Eqn:eff_1})-(\ref{Eqn:eff_4}) [blue curve in Fig. \ref{Fig:Broad_near_field_theory}(\textit{c})].
For comparison, also displayed are the peak locations of
the wave packets predicted by the ray-based theory (\ref{Eqn:ray_base})-(\ref{Eqn:wavenum_conserve}) [red curve in Fig. \ref{Fig:Broad_near_field_theory}(\textit{c})].
One can observe that there is
excellent agreement between the numerical results and theoretical predictions
for the motion of the peak locations.
This confirms that the wave
packet moves at the group velocity $v_{g}=\partial \overline{\nu }%
_{k}/\partial k$.
As predicted by the ray-based theory, for $0\leq T\leq 4300$, the group velocity
${v}_{g}\sim -0.045$ is negative with a relatively small magnitude, whereas for $4300\leq T\leq 4700$, the group velocity ${v}_{g}\sim 0.37$ is positive with a relatively large magnitude.
These two theoretical group velocities are in excellent agreement with the
measured ones, $\widetilde{v}_{g}\sim -0.045$ and
$\widetilde{v}_{g}\sim 0.37$, respectively.
As reflected in the zig-zag pattern in Fig. \ref{Fig:Broad_near_field_theory}(\textit{c}),
we can observe that SW packets propagate towards the
trailing edge with a relatively small group velocity, and towards the
leading edge with a relatively large group velocity.

(iii)
Finally, we discuss  the temporal evolution of the maximal
amplitude, $A$, of fast-mode waves in the near field.
Figure \ref%
{Fig:Broad_near_field_theory}(\textit{d}) displays the maximal amplitude of
the fast-mode waves in our TWN model (\ref{Eqn:eta1_moving})-(\ref%
{Eqn:M2_moving}) [green curve] and that predicted using the effective
linearized equations (\ref{Eqn:eff_1})-(\ref{Eqn:eff_4}) [blue curve]. One
can see very good agreement between them.
In addition, one
can observe from Fig. \ref{Fig:Broad_near_field_theory}(\textit{d})
that the amplitude $A$ is
relatively large for the negative group velocity $v_{g}$ (interval marked by the red color),
whereas the
amplitude $A$ is relatively small for the positive group velocity $v_{g}$ (interval marked by the black color).
Furthermore, for $4500\leq T\leq 9300$, the
amplitude $A$ grows for positive $X$ (interval marked by the magenta color),
whereas it decays for negative $X$ (interval marked by the cyan color).
Therefore, SWs become high in amplitude at the leading edge ($%
X>0$) whereas low at the trailing edge ($X<0$).

To the best of our knowledge, the asymmetric behavior {(i)} was earlier discovered in references %
\cite{Bakhanov2002JGR,Kodaira2016JFM}, the asymmetric behavior {(iii)} was earlier discovered in the reference \cite{Lewis1974JFM}, but the
asymmetric behavior {(ii)} was not reported in previous works. Here, we quantify these asymmetric types of behavior predicted by the
ray-based theory for our TWN model when the initial perturbation (\protect\ref{Eqn:perturb})
is a small-amplitude, narrow-width SW packet with a narrow band of wavenumbers.



\section{Conclusions and discussion}

\label{sec:conclusion}

%




Using the long-wavelength  and
small-amplitude approximations, we have proposed a two-layer, weakly nonlinear (TWN) model (\ref%
{Eqn:1.upkin})-(\ref{Eqn:1.dndyn}) describing the long-wave interactions between IWs and SWs.
The TWN model captures the broadening
of large-amplitude IWs that is a ubiquitous phenomenon in the ocean \cite{Perry1965JGR,Duda2004IEEEJOE}.
In Sec. \ref{sec:surface_wave}, we have investigated the spatiotemporal
manifestation and the dynamical behavior of right-moving fast-mode SWs in the near field in the presence of an underlying IW.
From our numerical results,  the wavenumber, group velocity, and
amplitude of fast-mode SW packets of our TWN model
(\ref{Eqn:eta1_moving})-(\ref{Eqn:M2_moving}) can always be well captured by
the predictions of the effective linearized equations
(\ref{Eqn:eff_1})-(\ref{Eqn:eff_4}) and the ray-based theory
(\ref{Eqn:ray_base})-(\ref{Eqn:wavenum_conserve}). The fast-mode waves behave as linear waves
modulated by the underlying interfacial solitary wave.
Importantly, the behavior of the right-moving fast-mode waves is asymmetric at the
leading edge vs. the trailing edge when an underlying IW is present:

\noindent (i) \label{asym2}\textit{SWs become short in wavelength at the
leading edge and long at the trailing edge [Fig. \ref{Fig:Broad_near_field_theory}(b)]. }

\noindent (ii) \label{asym1}\textit{SW packets propagate towards the
trailing edge with a relatively small group velocity, and towards the
leading edge with a relatively large group velocity [Fig. \ref%
{Fig:Broad_near_field_theory}(c)].}

\noindent (iii) \label{asym3}\textit{SWs become high in amplitude at the
leading edge and low at the trailing edge [Fig. \ref{Fig:Broad_near_field_theory}(d)]. }

In this work, we only focus on the spatiotemporal manifestation and dynamical behavior
of SWs under a small-amplitude initial perturbation.
As a natural extension of the above results, it is interesting to study the SWs when the amplitude of the perturbation is large, that is,  the nonlinearity becomes
prominent. In particular, it is important to understand
how the nonlinearity and resonance affect the spatiotemporal manifestation and dynamical behavior of the right-moving fast-mode SWs
in the presence of an underling IW.



Class 3 triad resonance is believed to be responsible for the
surface signature of the underlying internal waves \cite{Osborne1980Sci,Kropfli1999JGR,Hwung2009JFM,Craig2012JFM}. The TWN model possesses two-mode waves, one
slow and the other fast, and thus resonant interactions among different modes
can occur during the energy exchange process [more details can be found in Appendix \ref{c3_triad_resn}].
The class 3 triad resonance
condition [Eq. (\ref{Eqn:resonce_cond}) in Appendix \ref{c3_triad_resn}], $c_{g}(k)=C_{p}(0)$,  shows that, for resonantly-interacting
waves, the group velocity of fast-mode waves $c_{g}(k)$ and the phase velocity of slow-mode
waves $C_{p}(0)$ are equal \cite{Phillips1974ARFM,
Hashizume1980JPSJp}.
From many field observations, a narrow band of SWs with the resonant wavenumber was found to be located at the leading
edge of an underlying IW and travel nearly at  the same speed as the underlying IW \cite{Osborne1980Sci,Kropfli1999JGR}.
The surface phenomena may be related to both the triad resonance excitation and the three asymmetric types of behavior (i)-(iii).
The allowance of triad resonance in the TWN model encourages us to investigate the spatiotemporal manifestation and dynamical behavior of SWs under large-amplitude initial perturbations in future work.









\section{Acknowledgements}

This work is supported by NYU Abu Dhabi Institute G1301,
NSFC Grant No. 11671259, 11722107, and 91630208,  and SJTU-UM Collaborative
Research Program (D.Z.). We dedicate this paper to our late mentor
David Cai.


\appendix

\section{Effective linearized equations}

\label{effective_linear_eqn}

In this section, we present the effective linearized equations of
the TWN model (\ref{Eqn:eta1_moving})-(\ref%
{Eqn:M2_moving}).  The variables $(\xi _{i},\overline{u}_{i})$ in the TWN model are composed of two components,
one being the interfacial solitary wave $(\Xi _{i},\overline{U}_{i})$\ and the other the perturbation of fast-mode
waves $(\widehat{\xi }_{i},\widehat{\overline{u}}_{i})$. By substituting $%
(\xi _{i},\overline{u}_{i})=(\Xi _{i},\overline{U}_{i})+(\widehat{\xi }_{i},%
\widehat{\overline{u}}_{i})$ into the system (\ref{Eqn:eta1_moving})-(\ref%
{Eqn:M2_moving}) and collecting the linear terms with respect to $(\widehat{%
\xi }_{i},\widehat{\overline{u}}_{i})$, we can obtain the effective
linearized equations for the fast-mode waves as follows,
\begin{equation}
\widehat{\eta }_{1T}-C_{s}\widehat{\eta }_{1X}+\left[H_{1}\widehat{\overline{u}}%
_{1}+\overline{U}_{1}(\widehat{\xi }_{1}-\widehat{\xi }_{2})\right]_{X}=0,
\label{Eqn:eff_1}
\end{equation}%
\begin{equation}
\widehat{\eta }_{2T}-C_{s}\widehat{\eta }_{2X}+(H_{2}\widehat{\overline{u}}%
_{2}+\overline{U}_{2}\widehat{\xi }_{2})_{X}=0,  \label{Eqn:eff_2}
\end{equation}%
\begin{equation}
\widehat{M}_{1T}-C_{s}\widehat{M}_{1X}+\left( \overline{U}_{1}\widehat{%
\overline{u}}_{1}+g\widehat{\xi }_{1}\right) _{X}=0,  \label{Eqn:eff_3}
\end{equation}%
\begin{equation}
\widehat{M}_{2T}-C_{s}\widehat{M}_{2X}+\left( \overline{U}_{2}\widehat{%
\overline{u}}_{2}+g\widehat{\xi }_{2}+\rho _{r}g\widehat{\eta }_{1}\right)
_{X}=0,  \label{Eqn:eff_4}
\end{equation}%
where%
\begin{equation*}
H_{1}=h_{1}+\Xi _{1}-\Xi _{2},\text{ \ \ }H_{2}=h_{2}+\Xi _{2},
\end{equation*}%
\begin{equation*}
\widehat{\eta }_{1}=h_{1}+\widehat{\xi }_{1}-\widehat{\xi }_{2},\text{ \ \ }%
\widehat{\eta }_{2}=h_{2}+\widehat{\xi }_{2},
\end{equation*}%
\begin{equation*}
\widehat{M}_{1}=\widehat{\overline{u}}_{1}-\frac{1}{3}h_{1}^{2}\widehat{%
\overline{u}}_{1XX}-\frac{1}{2}h_{1}\left( H_{2}\widehat{\overline{u}}_{2}+%
\overline{U}_{2}\widehat{\xi }_{2}\right) _{XX},
\end{equation*}%
\begin{equation*}
\widehat{M}_{2}=\widehat{\overline{u}}_{2}-\frac{1}{2}\rho _{r}h_{1}^{2}%
\widehat{\overline{u}}_{1XX}-\frac{1}{3}h_{2}^{2}\widehat{\overline{u}}%
_{2XX}-\rho _{r}h_{1}\left( H_{2}\widehat{\overline{u}}_{2}+\overline{U}_{2}%
\widehat{\xi }_{2}\right) _{XX}.
\end{equation*}

\section{Class 3 triad resonance condition}

\label{c3_triad_resn}

\begin{figure}
\centering \hspace*{-2.0mm} \includegraphics[scale=0.7]{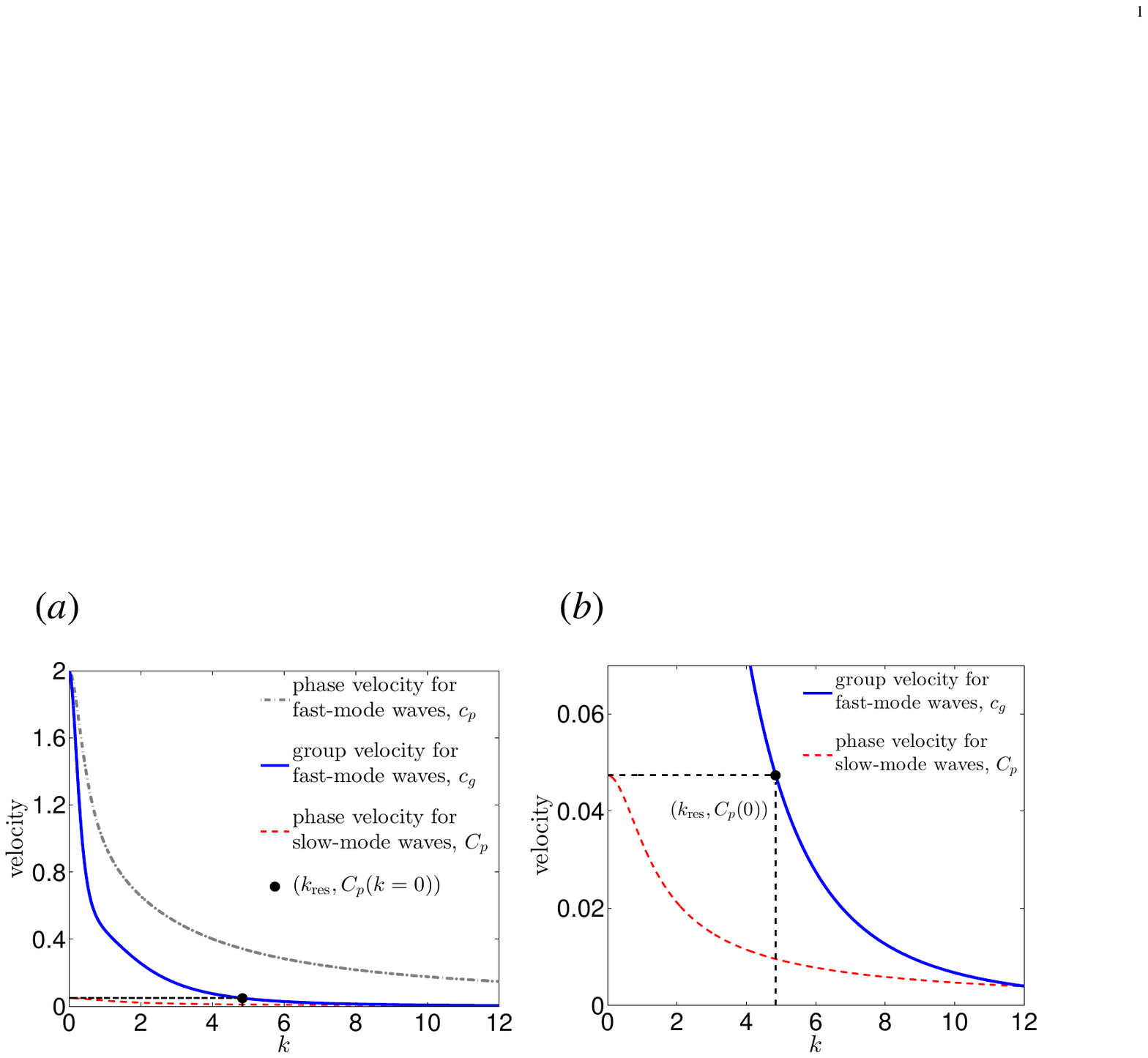}
\caption{(Color online) (\textit{a}) Comparison of the phase velocity for fast-mode waves $%
c_{p}$ , the group velocity for fast-mode waves $c_{g}$, and the phase velocity for
slow-mode waves $C_{p}$, with the parameters $(h_{1},h_{2},g,%
\protect\rho _{1},\protect\rho _{2})=(1,3,1,1,1.003)$.
The
resonant wavenumber $k_{\mathrm{res}}$ satisfies the triad resonance
condition (\protect\ref{Eqn:resonce_cond}) (\textit{b}) Zoomed-in version of panel (\textit{a}).
}
\label{Fig:resonant_exist}
\end{figure}

In this section, we briefly verify the existence of solutions to the
three-wave-resonance condition in the TWN model.  If
the dispersion relation allows the wavenumbers $k_{i}\ (i=1,2,3)$ and the
corresponding frequencies $\omega _{k}(k_{1})$, $\omega _{k}(k_{2})$, and $\Omega _{k}(k_{1})$  to satisfy the
conditions,
\begin{align*}
k_{1}&-k_{2} =k_{3}, \\
\omega _{k}(k_{1}) &- \omega _{k}(k_{2}) =\Omega _{k}(k_{3}),
\end{align*}%
these three waves constitute a class 3 resonant triad. Moreover, if the wavenumbers
are specified as  $k_{1}=k+\Delta k/2$, $k_{2}=k-\Delta k/2$, $%
k_{3}=\Delta k$, where $\Delta k\ll k$ and $\Delta k\rightarrow 0$, then the
resonance condition reduces to
\begin{equation}
c_{g}(k)=C_{p}(0),  \label{Eqn:resonce_cond}
\end{equation}%
where the group velocity $c_{g}$ and the phase velocity $C_{p}$ are given by
the equations
\begin{equation}
c_{g}(k)\equiv \frac{\partial \omega _{k}\left( k\right) }{\partial k},\ \ \ \
C_{p}(0)\equiv \left. \frac{\Omega _{k}(\Delta k)}{\Delta k}\right\vert
_{\Delta k\rightarrow 0}.  \label{Eqn:res_define}
\end{equation}%
Here, $\omega _{k}$ corresponds to the dispersion relation of fast-mode waves
in Eq. (\ref{Eqn:disp_fast}), and
$\Omega _{k}$ corresponds to the dispersion relation of slow-mode waves
in Eq. (\ref{Eqn:disp_slow}).
Many early results
\cite{Phillips1974ARFM,
Hashizume1980JPSJp,Craig2011NatHazd,Craig2012JFM,Alam2012JFM} have confirmed that there
exists a unique resonant wavenumber, denoted by $k_{\text{res}}$, satisfying Eq. (\ref{Eqn:resonce_cond}) in
the two-layer fluid system.
For the TWN model, one can observe from Fig. \ref{Fig:resonant_exist} that
there exists a unique resonant wavenumber $k_{\text{res}}$, satisfying the resonance
condition (\ref{Eqn:resonce_cond}), $c_{g}(k_{\text{res}})=C_{p}(k=0)$.
Therefore,
class 3 resonant triads exist among
two fast-mode waves and one slow-mode wave for the TWN model.

\bibliographystyle{plain}
\bibliography{Short_Format,myreference}

\begin{thebibliography}{10}

\bibitem{Alam2012JFM}
Mohammad-Reza Alam.
\newblock A new triad resonance between co-propagating surface and interfacial
  waves.
\newblock {\em J. Fluid Mech.}, 691:267--278, 2012.

\bibitem{Alam2009JFM}
Mohammad-Reza Alam, Yuming Liu, and Dick~KP Yue.
\newblock Bragg resonance of waves in a two-layer fluid propagating over bottom
  ripples. part i. perturbation analysis.
\newblock {\em J. Fluid Mech.}, 624:191--224, 2009.

\bibitem{Alford2015Nature}
M.~H. Alford, T.~Peacock, J.~A. MacKinnon, J.~D. Nash, M.~C. Buijsman, L.~R.
  Centuroni, S.-Y. Chao, M.-H. Chang, D.~M. Farmer, O.~B. Fringer, et~al.
\newblock The formation and fate of internal waves in the south china sea.
\newblock {\em Nature}, 521(7550):65--69, 2015.

\bibitem{Ambrosi2000WM}
D~Ambrosi.
\newblock Hamiltonian formulation for surface waves in a layered fluid.
\newblock {\em Wave motion}, 31(1):71--76, 2000.

\bibitem{Apel2007JASA}
J.~R. Apel, L.~A. Ostrovsky, Y.~A. Stepanyants, and J.~F. Lynch.
\newblock Internal solitons in the ocean and their effect on underwater sound.
\newblock {\em J. Acoust. Soc. Amer.}, 121(695¨C722), 2007.

\bibitem{Bakhanov2002JGR}
V.~V. Bakhanov and L.~A. Ostrovsky.
\newblock Action of strong internal solitary waves on surface waves.
\newblock {\em J. Geophys. Res.}, 107(3139), 2002.

\bibitem{Barros2009SAM}
R.~Barros and W.~Choi.
\newblock Inhibiting shear instability induced by large amplitude internal
  solitary waves in two-layer flows with a free surface.
\newblock {\em Stud. Appl. Math}, 122(325-346), 2009.

\bibitem{Barros2007SAM}
R.~Barros and S.~Gavrilyuk.
\newblock Dispersive nonlinear waves in two-layer flows with free surface part
  ii. large amplitude solitary waves embedded into the continuous spectrum.
\newblock {\em Stud. Appl. Math}, 119(213-251), 2007.

\bibitem{Caponi1988DTIC}
E.~A. Caponi, D.~R. Crawford, H.~C. Yuen, and P.~G. Saffman.
\newblock Modulation of radar backscatter from the ocean by a variable surface
  current.
\newblock Technical report, DTIC Document, 1988.

\bibitem{Chen2005Efficient}
Tong Chen.
\newblock {\em An efficient algorithm based on quadratic spline collocation and
  finite difference methods for parabolic partial differential equations}.
\newblock PhD thesis, University of Toronto, 2005.

\bibitem{Choi2009JFM}
W.~Choi, R.~Barros, and T.-C. Jo.
\newblock A regularized model for strongly nonlinear internal solitary waves.
\newblock {\em J. Fluid Mech.}, 629(73-85), 2009.

\bibitem{Choi1996JFM}
W.~Choi and R.~Camassa.
\newblock Weakly nonlinear internal waves in a two-fluid system.
\newblock {\em J. Fluid Mech.}, 313(83-103), 1996.

\bibitem{Choi1999JFM}
W.~Choi and R.~Camassa.
\newblock Fully nonlinear internal waves in a two-fluid system.
\newblock {\em J. Fluid Mech.}, 396(1-36), 1999.

\bibitem{Craig2004CRM}
W.~Craig, P.~Guyenne, and H.~Kalisch.
\newblock A new model for large amplitude long internal waves.
\newblock {\em C. R. Mecanique}, 332(525-530), 2004.

\bibitem{Craig2005CPAM}
W.~Craig, P.~Guyenne, and H.~Kalisch.
\newblock Hamiltonian long wave expansions for free surfaces and interfaces.
\newblock {\em Commun. Pure Appl. Maths}, 58(1587-1641), 2005.

\bibitem{Craig2011NatHazd}
W.~Craig, P.~Guyenne, and C.~Sulem.
\newblock Coupling between internal and surface waves.
\newblock {\em Nat. Hazards}, 57(617-642), 2011.

\bibitem{Craig2012JFM}
W.~Craig, P.~Guyenne, and C.~Sulem.
\newblock The surface signature of internal waves.
\newblock {\em J. Fluid Mech.}, 710(277-303), 2012.

\bibitem{Dias2001PhysD}
F.~Dias and A.~Il'ichev.
\newblock Interfacial waves with free-surface boundary conditions: an approach
  via a model equation.
\newblock {\em Physica D}, 150(278-300), 2001.

\bibitem{Donato1999JFM}
A.~N. Donato, D.~H. Peregrine, and J.~R. Stocker.
\newblock The focusing of surface waves by internal waves.
\newblock {\em J. Fluid Mech.}, 384:27--58, 1999.

\bibitem{Duda1998DTIC}
T.~F. Duda and D.~M. Farmer.
\newblock The 1998 {WHOI/IOS/ONR} {I}nternal {S}olitary {W}ave {W}orkshop:
  {C}ontributed {P}apers.
\newblock Technical report, DTIC Document, 1999.

\bibitem{Duda2004IEEEJOE}
T.~F. Duda, J.~F. Lynch, J.~D. Irish, R.~C. Beardsley, S.~R. Ramp, C.~S. Chiu,
  T.~Y. Tang, and Y.~J. Yang.
\newblock Internal tide and nonlinear wave behavior in the continental slope in
  the northern south china sea.
\newblock {\em IEEE J. Ocean. Eng.}, 29(1105-1131), 2004.

\bibitem{Fochesato2005PhysD}
C.~Fochesato, F.~Dias, and R.~Grimshaw.
\newblock Generalized solitary waves and fronts in coupled korteweg-de vries
  systems.
\newblock {\em Physica D}, 210(96-117), 2005.

\bibitem{Funakoshi1983JPSJp}
M.~Funakoshi and M.~Oikawa.
\newblock The resonant interaction between a long internal gravity wave and a
  surface gravity wave packet.
\newblock {\em J. Phys. Soc. Jpn}, 56(1982-1995), 1983.

\bibitem{Gargettt1972JFM}
A.~E. Gargett and B.~A. Hughes.
\newblock On the interaction of surface and internal waves.
\newblock {\em J. Fluid Mech.}, 52(01):179--191, 1972.

\bibitem{Guyenne2006CRM}
Philippe Guyenne.
\newblock Large-amplitude internal solitary waves in a two-fluid model.
\newblock {\em Comptes Rendus M{\'e}canique}, 334(6):341--346, 2006.

\bibitem{Han2007AMC}
H.~Han and Z.~Xu.
\newblock Numerical solitons of generalized korteweg¨cde vries equations.
\newblock {\em Appl. Math. Comput.}, 186(483-489), 2007.

\bibitem{Hashizume1980JPSJp}
Y.~Hashizume.
\newblock Interaction between short surface waves and long internal waves.
\newblock {\em J. Phys. Soc. Jpn}, 48(631-638), 1980.

\bibitem{Helfrich2006ARFM}
K.~R. Helfrich and W.~K. Melville.
\newblock Long nonlinear internal waves.
\newblock {\em Annu. Rev. Fluid Mech.}, 38(395-425), 2006.

\bibitem{Hwung2009JFM}
H.-H. Hwung, R.-Y. Yang, and I.~V. Shugan.
\newblock Exposure of internal waves on the sea surface.
\newblock {\em J. Fluid Mech.}, 626(1-20), 2009.

\bibitem{Jackson2004Atlas}
Christopher~R Jackson and J~Apel.
\newblock An atlas of internal solitary-like waves and their properties.
\newblock {\em Contract}, 14(03-C):0176, 2004.

\bibitem{Jo2008SAM}
T.-C. Jo and W.~Choi.
\newblock On stabilizing the strongly nonlinear internal wave model.
\newblock {\em Stud. Appl. Math}, 120(65-85), 2008.

\bibitem{Kawahara1975JPSJ}
Takuji Kawahara, Nobumasa Sugimoto, and Tsunehiko Kakutani.
\newblock Nonlinear interaction between short and long capillary-gravity waves.
\newblock {\em J. Phys. Soc. Jpn.}, 39(5):1379--1386, 1975.

\bibitem{Kodaira2016JFM}
Tsubasa Kodaira, Takuji Waseda, Motoyasu Miyata, and Wooyoung Choi.
\newblock Internal solitary waves in a two-fluid system with a free surface.
\newblock {\em J. Fluid Mech.}, 804:201--223, 2016.

\bibitem{Koop1981JFM}
C~Gary Koop and Gerald Butler.
\newblock An investigation of internal solitary waves in a two-fluid system.
\newblock {\em J. Fluid Mech.}, 112:225--251, 1981.

\bibitem{Kropfli1999JGR}
R.~A. Kropfli, L.~A. Ostrovski, T.~P. Stanton, E.~A. Skirta, A.~N. Keane, and
  V.~Irisov.
\newblock Relationships between strong internal waves in the coastal zone and
  their radar and radiometric signatures.
\newblock {\em J. Geophys. Res.}, 104(3133-3148), 1999.

\bibitem{Lee2007JKrPS}
K.-J. Lee, I.~V. Shugan, and J.-S. An.
\newblock On the interaction between surface and internal waves.
\newblock {\em J. Korean Phys. Soc.}, 51(616-622), 2007.

\bibitem{Lewis1974JFM}
J.~E. Lewis, B.~M. Lake, and D.~R.~S. Ko.
\newblock On the interaction of internal waves and surface gravity waves.
\newblock {\em J. Fluid Mech.}, 63(04):773--800, 1974.

\bibitem{Osborne1980Sci}
A.~R. Osborne and T.~L. Burch.
\newblock Internal solitons in the andaman sea.
\newblock {\em Science}, 208(451-460), 1980.

\bibitem{Parau2001JFM}
E.~Parau and F.~Dias.
\newblock Interfacial periodic waves of permanent form with free-surface
  boundary conditions.
\newblock {\em J. Fluid Mech.}, 437(325-336), 2001.

\bibitem{Perry1965JGR}
R.~B. Perry and G.~R. Schimke.
\newblock Large-amplitude internal waves observed off the northwest coast of
  sumatra.
\newblock {\em J. Geophys. Res.}, 70(10):2319--2324, 1965.

\bibitem{Phillips1974ARFM}
O.~M. Phillips.
\newblock Nonlinear dispersive waves.
\newblock {\em Annu. Rev. Fluid Mech.}, 6(93-110), 1974.

\bibitem{Sepulveda1987PhysFld}
N.~Sepulveda.
\newblock Solitary waves in the resonant phenomenon between a surface gravity
  wave packet and an internal gravity wave.
\newblock {\em Phys. Fluids}, 30(7), 1987.

\bibitem{Tanaka2015JFM}
M.~Tanaka and K.~Wakayama.
\newblock A numerical study on the energy transfer from surface waves to
  interfacial waves in a two-layer fluid system.
\newblock {\em J. Fluid Mech.}, 763:202--217, 2015.

\bibitem{Whitham1974Linear}
G.~B. Whitham.
\newblock {\em Linear and Nonlinear Waves}.
\newblock A Wiley-Interscience Publication, New York, 1974.

\end{thebibliography}

\end{document}